\documentclass[draft]{agujournal2019}
\usepackage{url} %this package should fix any errors with URLs in refs.
\usepackage{lineno}
\usepackage[inline]{trackchanges} %for better track changes. finalnew option will compile document with changes incorporated.
\usepackage{soul}
\draftfalse

\journalname{JGR: Space Physics}

\begin{document}

\title{Modeling Interplanetary Expansion and Deformation of CMEs with ANTEATR-PARADE I: Relative Contribution of Different Forces}
%\title{Physics-Driven Modeling of Interplanetary CME Expansion and Deformation I: Relative Contribution of Different Forces}

\authors{C. Kay\affil{1,2}, T. Nieves-Chinchilla\affil{1}}

\affiliation{1}{Heliophysics Science Division, NASA Goddard Space Flight Center, Greenbelt, MD, USA}
\affiliation{2}{Dept. of Physics, The Catholic University of America, Washington DC, USA}

\correspondingauthor{Christina Kay}{christina.d.kay@nasa.gov}

%  List up to three key points (at least one is required)
%  Key Points summarize the main points and conclusions of the article
%  Each must be 100 characters or less with no special characters or punctuation and must be complete sentences

\begin{keypoints}
\item ANTEATR-PARADE adds CME expansion and deformation to the interplanetary drag model
\item Drag and any form of cross sectional pressure gradient are the most important forces for deformation
\item The initial expansion velocity has a very strong effect on the total amount of deformation
\end{keypoints}

%\begin{keypoints}
%\item A group of anteaters is called a parade.
%\item I need to space out writing papers better.
%\item I prefer writing code.
%\end{keypoints}

%% \begin{abstract} starts the second page

\begin{abstract}
Coronal Mass Ejections (CMEs) are key drivers of space weather activity but most predictions have been limited to the expected arrival time of a CME, rather than the internal properties that affect the severity of an impact. Many properties, such as the magnetic field density and mass density, follow conservation laws and vary systematically with changes in the size of a CME. We present ANTEATR-PARADE, the newest version of the ANTEATR arrival time model, which now includes physics-driven changes in the size and shape of both the CME's central axis and its cross section. Internal magnetic and thermal and external drag forces affect the acceleration of the CME in different directions, inducing asymmetries between the radial and perpendicular directions. These improvements should lead to more realistic CME velocities, both bulk and expansion, sizes and shapes, and internal properties. We present the model details, an initial illustration of the general behavior, and a study of the relative importance of the different forces. The model shows a pancaking of both the cross section and central axis of the CME so that their radial extent becomes smaller than their extent in the perpendicular direction. We find that the initial velocities, drag, any form of cross section expansion, and the precise form of thermal expansion have strong effects. The results are less sensitive to axial forces and the specific form of the cross section expansion.
\end{abstract}

\section*{Plain Language Summary}
Coronal Mass Ejections (CMEs) are large explosions of matter and magnetic field that violently erupt from the Sun. When they hit the Earth they cause negative effects in human technology so it is important to forecast them. Most models of CMEs only predict if and when a CME could impact the Earth but not the properties that affect the severity of the impact. Many of these properties scale with the CME size and shape so we need to understand how the CME expands between the Sun and the Earth. We have taken a model for the arrival time and added new forces to better understand the evolution of a CME's size and shape. This should help us better predict factors like magnetic field strength, number density, and velocity. Our model reproduces a previously known trend of CMEs becoming squished in the radial direction or ``pancaking.'' We explore which forces are the most important for causing this effect. We find that drag forces and the expansion of the CME cross section are more important than axial forces. We also see that the model results depend on how we initially convert the total CME speed into initial propagation and expansion speeds.

\section{Introduction}
Understanding the interplanetary behavior of CMEs is critical for accurate space weather forecasting. The severity of a geomagnetic storm depends on the CME properties at the time of impact, mostly significantly the magnetic field but also its size and kinematic properties. Accordingly, we must not only understand the properties with which a CME is initiated, but how they evolve during its propagation if we wish to know the severity and timing of an impact \cite<e.g.>{Man17}. \citeA{Kil19} provides an in depth summary on many of the challenges of forecasting CMEs.

We tend to have an abundance of coronal images from the Earth’s perspective, both in visible light and the extreme ultraviolet, which have allowed us to study the source region and early evolution of Earth-impacting CMEs for nearly half of a century \cite<e.g.,>{Tou73}, but routine observations from off the Sun-Earth line are limited to the coronagraphs or heliospheric imagers on the STEREO satellites. Parker Solar Probe and Solar Orbiter will provide exciting remote observations from new angles and distances but will not give the consistent off-axis perspective needed to better understand the interplanetary evolution of CMEs. 

In contrast, in situ observations yield a single path through a CME, but it is often unclear where a satellite intersects a CME, or even which CME it is during times of high activity. Tying these observations together requires accurate modelling of a CME from the corona and through interplanetary space. Many case studies exist where large teams are able to work together to piece together the full Sun-to-Earth behavior of a single CME \cite<e.g.,>{Liu13,Liu16, Mos15, Pat16, Pal18,Hei19} but currently this is only done long after a CME's passage.

The vast majority of interplanetary CME models focus solely on predicting their arrival time at Earth \cite<e.g.,>{Vrs13,Mos15,Pao17,Liu18PUMA}, which is an essential goal for space weather forecasting. Arrival time models can be highly successful for forecasting but not lead to an improved understanding of the actual physics if they are empirical models fine-tuned to yield accurate results or highly-layered, machine-learning models that are opaque to their users.

The physics-driven drag models tend to treat the CME as a simplified structure and incorporate the effects of the standard drag equation in either one or two dimensions \cite<e.g.,>{Vrs13,Hes15,Mos15,Rol16,Nap18}. This is clearly an oversimplification of the actual physical processes at play. For example, \citeA{Owe17} argue that CMEs cannot be a coherent structure on large scales since their expansion speeds often exceed the local Alfven speed so information cannot be propagating over the full structure. These sort of toy models, however, run on the time scales needed for space weather prediction and have shown to be useful in predicting arrival times \cite<e.g.,>{Ril18,Wol18} while providing some level of physical intuition about the evolutionary processes.

Most of the models designed with predictions in mind focus solely on the arrival time of the CME, and possibly velocity, but do not provide a complete picture of the internal CME properties. \citeA{Kum96} developed a model for the evolution of CME flux rope using the Lorentz force and various conservation laws, giving one of the earliest, thorough descriptions of a CMEs coronal and interplanetary behavior. This model implies scaling laws for various CME properties with distance that compare favorably with interplantery observations of CMEs.  Recently, \citeA{Mis18} developed a model that simulates the internal thermodynamics of a CME during coronal propagation. This model uses the observed CME speeds in combination with Lorentz, thermal, and centrifugal forces to determine the relative importance of each as well as the thermodynamic properties. \citeA{Mis18} only present coronal results for a single observed CME but develop a method that could be useful for forward modelling many of the internal CME properties out to greater distances.  \citeA{Dur17} develop a model for interplanetary propagation of CMEs using a thermodynamic approach, but assume that the magnetic energy remains constant.

Interplanetary studies show that CMEs commonly become oblate during propagation so that their extent in the radial direction is much shorter than their extent in the direction perpendicular to the radial (hereafter perpendicular direction). This effect is often referred to as ``pancaking.'' Pancaking can be directly seen in simulations of various complexity \cite{Ril04, Ril04MHD, Sav11} and an oblate cross section can also be inferred from in situ observations of properties such as the shock standoff distance \cite{Rus02} or the direction flow in the sheath between the CME and shock \cite{Owe04}.

Previous studies have simulated the pancaking of a CME's cross section by assuming that all parts of it are convected out at the same speed in the local radial direction \cite{Ril04, Owe05,Liu06}. This naturally causes the CME to maintain a constant angular width, however, the distance between two points along the same radial direction will remain constant as they move at the same speed. This causes the aspect ratio to change since the cross section grows in the perpendicular direction while remaining fixed in the radial direction. 

Some models of the magnetic field of a CME can incorporate this ellipticity of the cross section \cite{Mul01,Hid02,Isa16,Nie18}. The majority of the elliptical magnetic field models are designed to be fit to observations by adjusting their free parameters in response to some sort of error minimization technique, rather than being used to forward model the evolution of a CME's magnetic field. \citeA{Isa16} illustrate the extent that evolutionary effects such as pancaking, expansion, and a skew of the central axis can have on the in situ profiles of a CME. \citeA{Nie18} introduce the elliptic-cylindrical analytical flux rope model, which is highly flexible in terms of describing a distorted flux rope cross section but simple enough to be of use for derivations that require the full expression of the magnetic field.

Similar to the distortion of the cross section shape, one may expect that the central axis of a CME also distorts during its interplanetary propagation. \citeA{Jan13} use in situ measurements to infer the local axis orientation of CME from 15 years of WIND observations at 1 AU and compare this with the expected distributions based on the global axis shape. They find that their results are compatible with the expected results for a CME that is 20\% wider in the perpendicular direction than it is in the radial direction, but not with a circular shape or greater than a 30\% asymmetry. A circular axis is often assumed for any models that are fit to coronal observations \cite<e.g. the Graduated Cylindrical Shell model,>{The06}, suggesting that a pancaking-like effect must occur in this direction in order to tie together the near-Sun and near-Earth measurements. An actual, detailed study of the coronal shapes of CMEs is required to truly understand what the average shape at 1 AU implies for the interplanetary evolution of the central axis.

We have developed a suite of strategically-simplified physics-driven models of CME behavior from the Sun to the Earth, with the intended eventual use for space weather predictions. The Open Solar Physics Rapid Ensemble Information (OSPREI) suite began with Forecasting a CME's Altered Trajectory \cite<ForeCAT,>{Kay15}, which modeled the deflection and rotation of a CME in the corona from background magnetic forces. We later developed the ForeCAT In situ Data Observer \cite<FIDO,>{Kay17FIDO} to combine a simple flux rope model with ForeCAT results to produce synthetic in situ profiles. These were then linked with ANother Type of Ensemble Arrival Time Results \cite<ANTEATR,>{Kay18, Kay20}, which determines transit times and velocities using a one-dimensional drag model but full three-dimensional CME shape when determining the precise timing of the impact. In this work, we combine ANTEATR with the magnetic field model of \citeA{Nie18} as ANTEATR-PARADE (Physics-driven Approach to Realistic Axis Deformation and Expansion) to develop the first forward model of a CME's interplanetary expansion and deformation that runs efficiently enough to be used for future real-time ensemble predictions. We present a basic description of the model in Section \ref{Model} and full details of the force derivations in the Supplementary Material. We then show the general model behavior for CMEs of different strengths and identify which forces are most responsible for the evolution of the CME parameters. A companion paper, \citeA{KayAP2} (hereafter Paper II), will perform a full sensitivity study of the model using parameter space explorations to determine the dependence of each output on each input.

\section{ANTEATR-PARADE Model}\label{Model}
The ANTEATR-PARADE models builds upon the original ANTEATR model by merging it with the analytical flux rope model of \citeA{Nie18}. Incorporating the internal magnetic forces of the CME allows us to simulate changes in the axis and cross section (CS) shape of a CME during interplanetary propagation. Here we explain the basic components of the model and describe the general algorithm. More details on nonorthogonal coordinate systems and magnetic forces in those coordinates can be found in the Supplementary Material and the full derivation of the magnetic field model can be found in \citeA{Nie18}.

\subsection{CME Shape}\label{shape}
\begin{figure}
 \noindent\includegraphics[width=\textwidth]{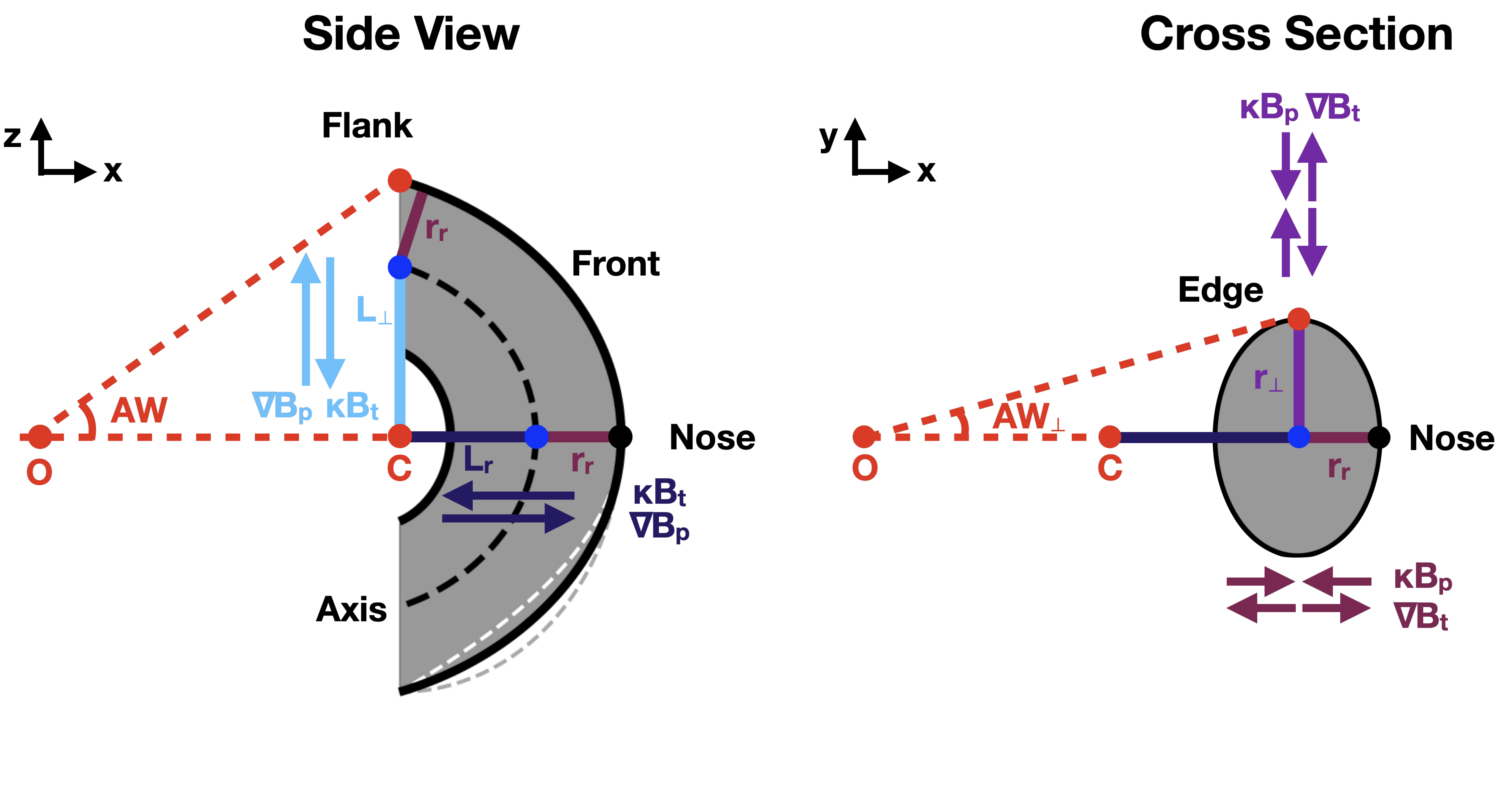}
\caption{Cartoon showing a side view and the cross section of the torus shape used in ANTEATR-PARADE.}
\label{cartoon}
\end{figure}

Previously, the OSPREI suite of models, including ANTEATR, has represented the flux rope of a CME as a torus with an elliptical axis and circular cross section. Figure \ref{cartoon} shows a side view of our new CME shape (left) and a view of the CS (right). In this cartoon, the $\hat{x}$ direction represents the radial direction at the CME nose, $\hat{y}$ represents the direction perpendicular to the radial vector in the plane of the CS at the nose, and $\hat{z}$ represents the direction perpendicular to the $\hat{x}$ in the plane containing the CME axis.

The first major change is to allow the CS to be an ellipse instead of restricting it to a circle, as shown on the right side of Fig. \ref{cartoon}, the same as in \citeA{Nie18}. The CS is now defined by the radius in the radial direction, $r_r$, and the perpendicular radius, $r_{\perp}$. We define the ratio of the radii as $\delta_{CS} = r_r/r_{\perp}$. The perimeter of the cross section is given by
\begin{eqnarray} \label{CSeqs}
  x  = \delta_{CS} r_{\perp} \cos \psi \nonumber \\
  y  = \hspace{16 pt} r_{\perp} \sin \psi
\end{eqnarray}
where $\psi$ is used for parameterization. The regions within the cross section are described by Equation \ref{CSeqs} with $r_{\perp}$ replaced by the coordinate $r$, which varies between 0 and $r_{\perp}$.  We emphasize that, unless $\delta_{CS}=1$, $r$ and $\psi$ are not equivalent to  a polar coordinate system with the origin at the center of the cross section. They form a nonorthogonal coordinate system and must be treated appropriately using covariant and contravariant coordinates and a tensorial analysis when calculating forces. The gist of the ANTEATR-PARADE model and certainly its results can be understood without these details so we only include them in the supplementary material and present a simplified description here. We caution the reader, however, on their own application of the equations in Section \ref{Model} without consulting the supplementary material. 

The second major change is in the shape of the toroidal axis. The toroidal axis is represented by the dashed black line in Fig. \ref{cartoon} and the blue dots lie along it in the direction of the nose and the flank.  We previously represented the axis as half of an ellipse defined by the lengths in the radial and perpendicular directions, now called $L_r$ and $L_{\perp}$, which extend from the center of the CME (marked with a red dot and a `C') to the blue dots on the toroidal axis. Analogous to the CS, we define $\delta_{Ax} = L_r / L_{\perp}$. An elliptical axis is then given by
\begin{eqnarray}
  x  = \delta_{Ax} L_{\perp} \cos \phi \nonumber \\
  z  = \hspace{16 pt} L_{\perp} \sin \phi
\end{eqnarray}
where $\phi$ is another ellipse parameterization and not a polar angle.

Typically, $L_r$ is the semi-minor axis and the toroidal axis is squashed in the radial direction relative to the perpendicular direction. For a full ellipse, the curvature at semi-major axis is greater than the curvature at the semi-minor axis. This corresponds to higher curvature at the CME flank than the CME nose. The precise behavior of a CME shape near the flanks is not yet fully understood but coronagraph images do not suggest it should be as tightly curved as a full ellipse. 

Alternatively, we can approximate the axis as the parabola that intersects the same two blue dots as the previous ellipse. This parabola is given by
\begin{equation}\label{parEq}
    x = \delta_{Ax} L_{\perp} - \frac{\delta_{Ax}}{L_{\perp}} z^2 
\end{equation}
This significantly increases the curvature at the nose while significantly decreasing it near the flank, likely beyond what is reasonable at either location. We form a hybrid shape, taking the average of the ellipse and the parabola to yield 
\begin{eqnarray}\label{hybrid}
  x  &=& \delta_{Ax} L_{\perp} \cos \phi \nonumber \hspace{66pt}\\
  z &=& \frac{1}{2} L_{\perp} \left(\sin\phi + \sqrt{1-\cos\phi} \right) 
\end{eqnarray}
which we derive by parameterizing Eq. \ref{parEq} with $\theta$ such that $z = L_{\perp}\sin\theta$ and relating $\phi$ and $\theta$ by finding where the two shapes give the same $x$ value. This causes a more gentle variation in the curvature with slightly stronger curvature at the nose than the edges.

The axial magnetic tension force we calculate (details in Section \ref{forces}) is quite sensitive to the curvature but these shapes are really quite similar visually. Fig. \ref{cartoon} shows the actual hybrid axis shape. The dashed white and grey lines interior and exterior to the CME front on the bottom of the side view in Fig. \ref{cartoon} show the change in the front for the parabola and ellipse axis, respectively. This is a very minor change in the apparent shape that cannot be constrained by current observations. Our use of the hybrid shape is fully motivated by it yielding the most stable forces during model development.

The toroidal axis defines the orientation of the CS as we define the CS to be perpendicular to the axis. At the nose this is the $xy$-plane, as shown in the  Fig. \ref{cartoon}, but it rotates toward the $yz$-plane as one moves along the axis. The normal direction to the axis can be calculated from the parametric definition of the axis. A general solution exists, but is quite convoluted. At the flank, the CS plane is not the $yz$-plane (note the orientation of the maroon line showing $r_r$ at the flank in Fig. \ref{cartoon}). Instead, we find that it is at an angle $\theta_n = \tan^{-1}(4 \delta_{Ax})$ with respect to the $x$-axis. For $\delta_{Ax}$ between 0.7 and 0.9 this corresponds to $\theta_n$ between 70.3$^{\circ}$ and 74.5$^{\circ}$, but for a $\delta_{Ax}$ of 0.25 it decreases to 45$^{\circ}$. 

Fully determining the CME shape requires the four lengths $r_r$, $r_{\perp}$, $L_r$, and $L_{\perp}$. Alternatively, we can define the radial distance of the front of the CME, $R_{F}$, the perpendicular or edge-on angular width, $AW_{\perp}$, and $\delta_{CS}$ to set $r_r$ and $r_{\perp}$. Finding $L_r$, and $L_{\perp}$ additionally requires $AW$ and $\delta_{Ax}$. We choose to use the combination of the $AW$s and $\delta$s to define our CME parameters as we find them more physically intuitive and potentially easier to measure in coronagraph images.

Since the normal vector typically has a $z$-component, the precise flank location corresponds to an axis position at some $\phi=\phi_{crit}$ greater than 90$^{\circ}$. This makes exactly calculating the $L$ and $r$ lengths from the angular widths and $\delta$s analytically intractable. Numerical solutions exist but greatly increase the computation time. We define the angular width as $\tan AW = R_E / R_C$ where $R_C$ is the distance from the center of the Sun to the center of the CME (red dot `O' to `C' in Fig. \ref{cartoon}) and $R_E$ is the distance from the CME center to the edge. The precise $R_E$ would be $L_{\perp}\sin(\phi_{crit})+r_r\hat{n}(\phi_{crit})$ but we simplify this to $L_{\perp}+r_r$. Mathematically this is equivalent to a slight deformation of our analytic CME front near the flanks but no change to the axis shape. Real CMEs do not adhere to a precise analytic form and the approximation produces negligible changes in our results so we proceed with this simplification.

\subsection{Magnetic Field}\label{Bmod}
ANTEATR-PARADE uses the elliptic-cylindrical flux rope model from \citeA{Nie18} (hereafter, NC18) for the CME's magnetic field. This model does not impose any internal force requirements, such as being force-free, which will allow us to explore different force distributions leading to distortions in the CME shape. NC18 solve the Maxwell equations for a generalized expression of the magnetic field in the nonorthongonal elliptical coordinate system. NC18 assume that there is no radial component of the magnetic field in the direction normal to the ellipse and that there are no changes in values along the cylindrical axis (analogous to our toroidal axis). By expressing the components of the current density as polynomials that depend only on $r$, NC18 derive an expression for the magnetic field components
\begin{eqnarray}\label{fullBeq}
  B_r &=& 0 ,\nonumber \\
  B_t &=&\sum_{n=1}^{\infty} \delta_{CS} B_n^0 [\tau - \bar{r}^{n+1}], \nonumber \\
  B_p &=& -\sum_{m=0}^{\infty}  \frac{(n+1)\delta_{CS}\sqrt{\delta_{CS}^2 \sin^2 \psi + cos^2\psi}}{\delta_{CS}^2 +m + 1} \frac{B_n^0}{C_{nm}} \bar{r}^{m+1},
\end{eqnarray}
where the NC18 $B^y$ and $B^{\psi}$ components correspond directly to our toroidal field, $B_t$, and poloidal field, $B_p$. In Equation \ref{fullBeq}, $n$ and $m$ are the order of the polynomial components. $\tau$ determines the internal twist distribution and with the scaling factor $C_{nm}$, determines the ratio of the toroidal and poloidal magnetic field. $B_n^0$ uniformly scales both the toroidal and poloidal magnetic field strength and $\bar{r}$ is the normalized radial distance that varies between 0 at the center and 1 at the edge. The center of the flux rope has a magnetic field strength of $\delta_{CS} \tau B_n^0$.

As in NC18, we restrict our magnetic field to [$m$,$n$] = [0,1] so that the expressions for the magnetic field becomes
\begin{eqnarray}\label{Beq}
  B_r &=& 0, \nonumber \\
  B_t &=& \delta_{CS} B_0 [\tau - \bar{r}^{2}], \nonumber \\
  B_p &=& -\frac{2 \delta_{CS} \sqrt{\delta_{CS}^2 \sin^2 \psi + cos^2\psi}}{1+\delta_{CS}^2} \frac{B_0}{C} \bar{r},
\end{eqnarray}
where we have replaced $B_1^0$ and $C_{10}$ with $B_0$ and $C$ for readability since we are only considering single values for $m$, $n$, $C$, and $\tau$. Currently, we do not have a physical understanding of the internal structure of CME flux ropes and cannot use such to constrain these parameters. \cite{Flo20} explores different combinations of these parameters to determine which lead to stable flux ropes. In Paper II we determine the variations with $C$ and $\tau$ and future work will explore the effects of different values of [$m$,$n$] on the ANTEATR-PARADE model.

\subsection{Magnetic Forces}\label{forces}
ANTEATR-PARADE calculates magnetic forces which act to both expand and deform the CME axis and CS. We calculate the magnetic tension and magnetic pressure gradients from both $B_t$ and $B_p$. The colored arrows in Fig. \ref{cartoon} show the direction of these forces relative to the CME shapes. For the CS, the poloidal magnetic tension (labeled $\kappa B_p$) will cause inward constriction whereas the toroidal magnetic gradient will cause outward expansion (labeled $\nabla B_t$) and are functions of $\psi$, in general, and can vary between the nose (maroon arrows) and the edge (purple arrows). The toroidal magnetic tension (labeled $\kappa B_t$) from the curved toroidal axis will cause it to move inward whereas the poloidal pressure gradient or hoop force (labeled $\nabla B_p$) will cause outward motion. As with the CS, we may not have a balance between the axis force at the nose (blue arrows) and the flank (light blue arrows). For both the CS and toroidal axis, an imbalance in magnetic tension and pressure will cause expansion, changing $AW$ and $AW_{\perp}$, and if the expansion is not balanced in different directions then $\delta_{CS}$ and $\delta_{Ax}$ will also change.

Determining these forces requires careful analysis in the nonorthogonal coordinate system. Here we present an overview of the process, full details are in the supplementary material. We start with the expression for the Lorentz force in terms of the current density and magnetic field from NC18 (Eq. 27). This equation is for a cylinder, rather than a curved tube, and the force points entirely in the $r$ direction, which is the normal to the ellipse at that particular $\psi$. We consider the forces from the axial curvature separately. We then use Ampere's law (Eqs. 13 and 14 in NC18) to replace the current density in the Lorentz force with derivatives of the magnetic field. This expression can be rearranged to contain terms analogous to the magnetic tension and magnetic pressure gradient forces one finds in an orthogonal coordinate system. These forces act on the CS with the poloidal tension constricting it and the toroidal gradient causing expansion, assuming that it exceeds the inward pressure gradient of the solar wind that we also include. For our chosen values of [$m$,$n$] = [0,1] the poloidal pressure gradient terms go to zero.

We consider a thin segment of the toroidal axis of width $R_{\kappa}d\phi$, where $R_{\kappa}$ is the local radius of curvature of the axis, and a wedge of width $rd\psi$ within this segment. The total force on this wedge is the Lorentz force per volume integrated over $r$, which we set equal to an acceleration multiplied by the density and the volume of the wedge. This gives us the acceleration of the edge of CS, $a_{CS}$.
\begin{equation}\label{aCS}
    a_{CS} = \frac{\delta_{CS}^2 B_0^2}{\pi \rho r_{\perp}}\left[ \frac{2}{3(1+\delta_{CS}^2)C^2} - \left(\frac{1}{3}\tau - \frac{1}{5} \right)\right] - \frac{B_{SW}^2}{8\pi r_{\perp}}
\end{equation}
This expression includes the inward pressure from the external solar wind magnetic field, unlike the version in the supplementary material which only includes the internal magnetic forces. This is the change in the normalized parametric $r$ and it does not depend on $\psi$, meaning that while these forces can cause the CS to grow to contract, the actual shape, defined by $\delta_{CS}$, will not change. This is specific to our chosen [$m$,$n$] = [0,1] and not necessarily true for any other combination.

The previous forces apply to a straight cylindrical or a curved flux rope but a curved central axis will also have an inward tension force from toroidal magnetic field and an outward hoop force from the poloidal field. To conserve magnetic flux, the poloidal field is enhanced on the side of the CS toward the center of curvature due to the decrease in area from the axial curvature. The opposite occurs on the side of the CS near the front of the CME, creating a outward gradient force. \citeA{Wel18} present a derivation of the hoop force in relation to the low coronal dynamics of CMEs for a CME with circular CS and circular axis. \citeA{Wel18} consider a segment of the torus with width $Rd\phi$, where $R$ is the radius of their circle (and the radius of curvature). We take a similar approach but consider a segment $R_c d \phi$ where $R_c$ is the local radius of curvature for the toroidal axis at some $\phi$. The magnitude of the toroidal field is unchanged but the poloidal field scales as 
\begin{equation}
    B_p'= B_p \frac{R_c}{R_c + \delta_{CS}r_{\perp}\bar{r}\cos\psi}
\end{equation}
where $B_p'$ is the poloidal field for a curved toroidal axis.

We approximate the segment as locally elliptic-cylindrical and use the poloidal pressure gradient terms of the Lorentz force in elliptic-cylindrical coordinates, which no longer go to zero. This leads to an acceleration of the toroidal axis of 
\begin{equation}
    a_{hoop} = \frac{B_0^2}{\pi \rho C^2 R_c} \frac{\sqrt{1-\delta_{CS}^2\gamma^2} (\delta_{CS}^2 \gamma^2 -6) + 6 - 4 \delta_{CS}^2 \gamma^2}{\delta_{CS}^3 \gamma^2 (1+\delta_{CS}^2)^2  \sqrt{1-\delta_{CS}^2\gamma^2}}
\end{equation}
where $\gamma=r_{\perp}/R_C(\phi)$ has been used to simply the expression.

For the axial magnetic tension we again consider a segment $R_c d \phi$ and use a curvature of $\kappa = 1/R_c$. The acceleration from the axial tension force  is
\begin{equation}
    a_{\kappa Bt} = \frac{\delta_{CS}B_0^2}{4\pi \rho R_c} \left( \tau^2 - \tau +\frac{1}{3}\right)
\end{equation}
which has a dependence on $\phi$ through the radius of curvature. Both the axial tension and hoop forces have a dependence on $\phi$ so the axial magnetic forces can change the shape and $\delta_{Ax}$ as well as causing expansion or contraction. Note that both these accelerations point in the direction normal to the toroidal axis. At the nose the normal direction is parallel to $\hat{x}$ so the forces fully contribute to expanding or contracting $L_r$. At the flank the normal is not parallel to $\hat{z}$ so $\sin(\theta_n)$ of the acceleration for $\phi=$90$^{\circ}$ affects $L_{\perp}$ but $\cos(\theta_n)$ of it affects $L_r$.

The magnetic forces give the expansion or contraction of the toroidal axis and CS, encompassing the internal magnetic forces. While the above equations hold over all $\phi$ and $\psi$ we only use the values at the nose and edge for the axis and similarly only along the two axes of the CS. From the magnetic forces, we determine the change in the four length parameters defining the CME shape.
\begin{eqnarray}
  \frac{\partial^2 r_r}{\partial t^2} &=& \delta_{CS} a_{CS} \nonumber \\
  \frac{\partial^2 r_{\perp}}{\partial t^2} &=& a_{CS} \nonumber \\
  \frac{\partial^2 L_r}{\partial t^2} &=& a_{hoop,n} + a_{\kappa Bt,n} - \left(a_{hoop,f} + a_{\kappa Bt,f}\right) \cos \theta_n   \nonumber \\
  \frac{\partial^2 L_{\perp}}{\partial t^2} &=& \left(a_{hoop,f} + a_{\kappa Bt,f}\right) \sin \theta_n 
\end{eqnarray}
where the last part of the subscript indicates axial accelerations at either the nose or the flank. We subtract the component of the flank acceleration in the $x$ direction as it corresponds to a change in the position of the sunward side of $L_r$.

\subsection{Drag Forces}\label{drag}
ANTEATR-PARADE also includes the external drag on the CME. We use the standard hydrodynamic drag equation as was previously done in the original ANTEATR. The force from drag is
\begin{equation}
    F_{drag} = - C_d A \rho_{SW} (v - v_{SW}) |v - v_{SW}|
\end{equation}
where $C_d$ is the dimensionless drag coefficient, $A$ is the cross-sectional area in plane perpendicular to the direction of motion, $v$ is a CME velocity, and $v_{SW}$ is the solar wind velocity, which we assume flows entirely in the radial direction. We determine the drag in three different directions. The first is the radial direction (the $x$-direction in Fig. \ref{cartoon}) 
\begin{equation}
    F_{d,r} = - C_d A_{FO} \rho_{SW} (v_F - v_{SW}) |v_F - v_{SW}|
\end{equation}
where $v_F$ is the velocity of the front, which is a combination of the bulk velocity, $v_B$, the axial expansion in the radial direction $v_{Ax,r}$, and the CS expansion in the radial direction, $v_{CS,r}$.
\begin{equation}
    v_F = v_B + v_{Ax,r} + v_{CS,r}
\end{equation}
The area perpendicular to the radial direction, $A_{FO}$, is the same as the face-on area, and can be determined from $AW$ and $AW_{\perp}$. 

The second drag force is in direction of the flanks (the $z$-direction in Fig. \ref{cartoon}), affecting the expansion of $AW$. 
\begin{equation}
    F_{d,\perp} = - C_d A_{\perp} \rho_{SW} (v_{F\perp} - \sin AW v_{SW}) |v_{F\perp} - \sin AW v_{SW}|
\end{equation}
Here, $v_{F\perp}$ is the velocity of the flank, which results from the perpendicular axial expansion, $v_{Ax,r}$, and the component of $v_{CS,r}$ in the $z$-direction.
\begin{equation}
    v_{F,\perp} = v_{Ax,\perp} + v_{CS,r} \sin \theta_n
\end{equation}
The cross-sectional area for this drag, $A_{\perp}$, can be determined from $AW_{\perp}$ and the length of the CME in the radial direction. 
The final drag force acts on the CS expansion in the perpendicular direction (the $y$-direction in Fig. \ref{cartoon}) and the CME velocity is simply the perpendicular CS expansion velocity, $ v_{CS,\perp}$.
\begin{equation}
    F_{d,CS,\perp} = - C_d A_{EO} \rho_{SW} (v_{F\perp} - \sin AW_{\perp} v_{SW}) |v_{F\perp} - \sin AW_{\perp} v_{SW}|
\end{equation}
Now the cross sectional area $A_{EO}$ is the same as the edge-on width and can be determined from the toroidal axis length and the radial width of the CS.

The first two drag forces are calculated using $v_F$ and $v_{F,\perp}$ and will clearly cause a change in these velocities but it is less obvious how they affect the individual velocities that make up $v_F$ and $v_{F,\perp}$. The net acceleration of the individual components should add up to the total acceleration. We (somewhat arbitrarily) decide to weight the total drag force by the fractional magnitude of the individual components relative to the total velocity
\begin{equation}
    F_{d,r} = \frac{v_B}{v_F} F_{d,r} +  \frac{v_{Ax,r}}{v_F} F_{d,r} +  \frac{v_{CS,r}}{v_F} F_{d,r} = F_{d,B} + F_{d,Ax,r} + F_{d,CS,r}
\end{equation}
where $F_{d,B}$, $F_{d,Ax,r}$, and $F_{d,CS,r}$ are the accelerations affecting the bulk, radial axial, and radial CS velocities. $v_{CS,r}$ appears in both the expression for $F_{d,r}$ and $F_{d,\perp}$. We assume that the $F_{d,CS,r}$ found at the nose is the same at the flank, which is an oversimplification for a real CME but allows us to retain a uniform CME CS. We subtract $\sin \theta_n F_{d,CS,r}$ from $F_{d,\perp}$ so that the remaining force is the drag on the axis in the perpendicular direction $F_{d,Ax,\perp}$

We note that we express our model in terms of drag forces instead of ram pressures, which is somewhat unconventional in the nonradial direction. Intrinsically, the drag force and ram pressure should be related, but much of the details get swept up into the drag coefficient. Since ANTEATR-PARADE has evolved out of an arrival time model, we embrace the drag in both directions but acknowledge this not being the standard approach.

\subsection{Thermal Forces}\label{thermF}
We incorporate a simple analytic model for the thermal pressure gradient causing additional expansion of the cross section. We assume a uniform temperature within the CME, $T_{CME}$, which is a clear oversimplification of the actual structure. In situ observations show significant fluctuations in the internal temperature of flux ropes \cite<e.g. cases in review by>{Man17} but we have not have an analytic model for any consistent internal structure. Currently, we initiate our uniform internal temperature by scaling it as some factor of the external solar wind temperature at the initial distance of the center of the CME cross section at the nose. Future work will incorporate more sophisticated CME temperature profiles once available. For the background solar wind temperature, we use 
\begin{equation}\label{Teq}
    T_{SW} \approx 62,000 \left(\frac{R}{1 \; \mathrm{AU}}\right)^{-0.58}
\end{equation}
from \citeA{Hel13}, where $R$ is the distance in AU. This empirical form was derived from Helios 1 and 2 measurements of the average proton temperature between 0.3 and 1 AU. We approximate the thermal pressure gradient using the internal CME pressure, the external solar wind pressure, and the cross-sectional radii $r_r$ and $r_{\perp}$. The internal CME pressure is $n_{CME} k T_{CME}$, where $n_{CME}$ is the CME number density and $k$ is the Boltzmann constant. The external solar wind pressure is $2n_{SW} k T_{SW}$. This leads to forces of 
\begin{eqnarray}
  \rho a_{T} = \rho \frac{\partial^2 r_r}{\partial t^2} &=& \frac{2k(n_{CME} T_{CME} - <n_{SW} T_{SW}>)}{ r_r} \nonumber \\
  \rho a_{T \perp} = \rho \frac{\partial^2 r_{\perp}}{\partial t^2} &=& \frac{2k(n_{CME} T_{CME} - n_{SW,E} T_{SW,E})}{ r_{\perp}}
\end{eqnarray}
where we assume the proton and electron pressure contribute equally for both the CME and the solar wind. We know that these temperatures are not equal at 1 AU within flux but do not have a better model for their interplanetary evolution so we start with the simplest approach and will improve the model as possible in the future. $<n_{SW} T_{SW}>$ represents the average of the $n_{SW} T_{SW}$ between the front and back of the cross section and $n_{SW,E} T_{SW,E}$ represent the values at the edge of the cross section. 

The exact thermodynamic nature of CME expansion is not fully understood. Statistical studies suggest that CMEs cool less rapidly than the background solar wind and that this process must be closer to isothermal than adiabatic \cite<e.g>{Wan05}, though recent models suggest an adiabatic approach may be reasonable \citeA{Dur17}. Using that $pV^{\gamma}$ is constant, where the thermal pressure, $p$, is proportional to the temperature divided by the volume, $V$, we find that $T V^{\gamma-1}$ is constant, where $\gamma$ is the adiabatic index. We leave $\gamma$ as a free parameter that can vary between isothermal ($\gamma$=1) and adiabatic ($\gamma$=5/3).

\subsection{Initial Velocity}\label{v0}
Typically, we begin a simulation at 10 $R_s$ as this is roughly where the external magnetic forces of the corona tend to become negligible. ANTEATR-PARADE is designed to take output from ForeCAT, our coronal deflection and rotation model, which we typically run out to 10 $R_s$. The internal forces may very well be important below this height, future work will incorporate them into the ForeCAT model and Paper II will study the sensitivity of ANTEATR-PARADE to small changes in $R_F$. 

This means that our initial parameters should describe the CME at 10 $R_s$. We only require the initial velocity of the CME front, which is often all that is easily measurable from a coronagraph image, which then needs to separated into $v_{CS,r}$, $v_{CS,\perp}$, $v_{Ax,r}$, $v_{Ax,\perp}$, and $v_B$, which we refer to as the initial velocity decomposition (IVD). The first approach is to assume that the CME is undergoing convective pancaking and extrapolate the approach of \citeA{Ril04} and \citeA{Owe05} to the full 3D torus structure. If the CME front is moving at $v_F$ and we assume constant angular widths $AW$ and $AW_{\perp}$ then the lengths defining the CME shape change as
\begin{eqnarray}
  r_{r} &=& r_{r,0} + v_{CS,r} t \; \; \; =  r_{r,0} + v_F (1 - \cos AW_{\perp}) t \nonumber \\
  r_{\perp} &=& r_{\perp,0} + v_{CS,\perp} t \; =  r_{\perp,0} + v_F \sin AW_{\perp} t \nonumber \\
  L_r &=& L_{r,0} + v_{Ax,r} t \; \; = L_{r,0} + v_F(\cos AW - \cos AW_{\perp}) t \nonumber \\
  L_{\perp} &=& L_{\perp,0} +  v_{Ax,\perp} t = L_{\perp,0} + v_F \left(\sin AW - \frac{1 - \cos AW_{\perp}}{\sin \theta_n}\right)t
\end{eqnarray}
where the subscript $0$ indicates initial values. Unlike \citeA{Owe05} we do not include any internal expansion within the convective velocity model since it is incorporated through our magnetic forces.

Alternatively, we can assume that the CME is initially fully self-similar and that both the angular widths, $AW$ and $AW_{\perp}$, and aspect ratios, $\delta_{CS}$ and $\delta_{Ax}$ remain constant in the absence of any forces. The front of the CME is initially at $R_{F0} = R_{C0} + L_{r0} + r_{r0}$, where $R_{C0}$ is the initial radial distance of the center of the CME (labeled with a $C$ in Fig. \ref{cartoon}) and not a radius of curvature. The lengths evolve kinematically, the same as for the convective velocities, but now the initial velocities are
\begin{eqnarray}
  v_{CS,r} &=& v_F \frac{r_{r0}}{R_{F0}} \nonumber \\ 
  v_{CS,\perp} &=& v_F \frac{r_{\perp 0}}{R_{F0}} \nonumber \\ 
  v_{Ax,r} &=& v_F \frac{L_{r0}}{R_{F0}} \nonumber \\ 
  v_{Ax,\perp} &=& v_F \frac{L_{\perp 0}}{R_{F0}}.  
\end{eqnarray}
For both the convective and self-similar approach, we use the model only to set the initial CME velocities and beyond this all velocities evolve according to the forces acting upon the CME.

These two options likely represent the two extremes of the possible values for a real CME, the ``correct'' values are probably not either, but somewhere in between. For now we do not have a definitive value from observations, and finding one from image analysis is a major undertaking, if at all possible, and far beyond the scope of this work. Here we consider both options for decomposing the front velocity into component velocities. In Paper II we present a method for varying between fully convective and fully self-similar and analyze the effect of the IVD on the outputs.

\subsection{ANTEATR-PARADE Algorithm}\label{algo}
We now have all the components necessary to build the basic algorithm of ANTEATR-PARADE. The model requires the initial speed of the CME front, $v_{F}$, the CME mass, $M_{CME}$, $AW$, $AW_{\perp}$, $\delta_{CS}$, $\delta_{Ax}$, the CME magnetic field strength and temperature relative to the background solar wind, and the properties of background solar wind at 1 AU. We first determine the initial values of $r_r$, $r_{\perp}$, $L_r$, and $L_{\perp}$. We then determine corresponding expansion velocities, $v_{CS,r}$, $v_{CS,\perp}$, $v_{Ax,r}$, and $v_{Ax,\perp}$ according to the choice of initial velocity model (convective or self-similar). 

Then, for each time step $\Delta t$, the magnetic, thermal, and drag forces are determined and parameters updated as follows
\begin{eqnarray}
  \Delta r_r &=&  v_{CS,r} \Delta t + \frac{1}{2}(\delta_{CS} a_{CS} +a_{d,CS,r}+ a_{T})\Delta t^2 \nonumber \\
  \Delta r_{\perp} &=&  v_{CS,\perp} \Delta t + \frac{1}{2}(a_{CS,\perp} +a_{d,CS,\perp} + a_{T\perp})\Delta t^2 \nonumber \\
  \Delta L_r &=& v_{Ax,r} \Delta t + \frac{1}{2} \left[\left(a_{hoop,f} + a_{\kappa Bt,f}\right) - \cos\theta_n \left(a_{hoop,f} + a_{\kappa Bt,f}\right) + a_{d,Ax,r} \right] \Delta t^2\nonumber \\
  \Delta L_{\perp} &=& v_{Ax,\perp} \Delta t +  \frac{1}{2} \left[\sin\theta_n\left(a_{hoop,f} + a_{\kappa Bt,f}\right) + a_{d,Ax,\perp} \right] \Delta t^2 \nonumber \\
  \Delta R_{F}  &=& v_F \Delta t + \frac{1}{2}\left[ \delta a_{CS} + \left(a_{hoop,f} + a_{\kappa Bt,f}\right)  + a_{d,r} \right] \Delta t^2 \nonumber \\
  \Delta v_{CS,r} &=& (\delta_{CS} a_{CS} +a_{d,CS,r}+ a_{T})\Delta t \nonumber \\
  \Delta v_{CS,\perp} &=& (a_{CS,\perp} +a_{d,CS,\perp}+ a_{T\perp})\Delta t \nonumber \\
  \Delta v_{Ax,r} &=& \left[\left(a_{hoop,f} + a_{\kappa Bt,f}\right) - \cos\theta_n \left(a_{hoop,f} + a_{\kappa Bt,f}\right) + a_{d,Ax,r} \right] \Delta t \nonumber \\
  \Delta v_{Ax,\perp} &=& \left[\sin\theta_n\left(a_{hoop,f} + a_{\kappa Bt,f}\right) + a_{d,Ax,\perp} \right] \Delta t \nonumber \\
  \Delta v_{F} &=& \left[ \delta a_{CS} + \left(a_{hoop,f} + a_{\kappa Bt,f}\right)  + a_{d,r} \right] \Delta t
\end{eqnarray}
where the drag accelerations, $a_{d,i}$, are determined from the corresponding forces $F_{d,i}$ by dividing by the CME mass. The CME density and magnetic field are then updated by assuming mass and magnetic flux conservation. We have two relations from the fluxes from $B_t$ and $B_p$ but three variables than can evolve to maintain flux conservation and are not already determined elsewhere in the model. The total magnetic field strength, $B_0$ is an obvious choice to have evolve but the system is underdetermined to calculate both $\tau$ and $C$. For now, we assume that $\tau$ remains constant and $C$ changes, which corresponds to changes in the ratio of the current densities and could imply a change in the stability of the flux rope. \citeA{Flo20} find that only certain combinations of $C$ and $\tau$ are stable with respect to the kink instability. ANTEATR-PARADE does not calculate any torques that could alter the orientation of the flux rope so approaching an unstable combination of $C$ and $\tau$ will not produce any kinking. We will, however, comment on model runs that cause $C$ and $\tau$ to evolve toward unstable values and therefore unviable results.

A simulation runs until the CME nose reaches a user-specified final distance. For now, all impacts occur at the CME nose (both $\phi$ and $\psi$ of 0$^{\circ}$). Future work will explore the effect of oblique impacts on the expected CME observables and fully develop the generation of in situ profiles.

For the background solar wind, we currently require the 1 AU values of the solar wind density, velocity, and total magnetic field strength, which we use to scale values at closer distances to the Sun. The solar wind velocity treated as constant and the density scales inversely with the distance squared. For the magnetic field we use a simple Parker spiral model.

\section{Ensemble Study Description}
Since these are the first results from the ANTEATR-PARADE model, we wish to understand the general behavior produced by the model more than the sensitivity to its parameters, which is the focus of Paper II. Here our focus is on understanding the relative importance of the different forces (drag, magnetic, and thermal) as well as understanding the actual CME evolution. We start with the most basic model possible, no forces at all, so that the CME evolves fully according to the IVD model. Then we add in drag, followed by the individual components of the magnetic forces (CS pressure only, CS tension, axial forces), and finally the thermal forces. For this process, we use an IVD corresponding to the average of the fully convective or full similar velocities. Then, for only the case with full forces, we also run the fully convective or fully self-similar cases to illustrate the effects of the IVD. Finally, we also run cases with either fully isothermal or adiabatic expansion.

We are left with the choice of input parameters, both the six that define the CME (v$_{F}$, $M_{CME}$, $AW$, $AW_{\perp}$, $\delta_{CS}$, $\delta_{Ax}$) and the magnetic field and thermal scale factors ($\beta_B = B_{CME} / B_{SW}$, $\beta_T = B_{CME} / B_{SW}$). We consider three different scale CMEs a slightly faster than average CME (which we refer to as average for simplicity hereafter), a fast CME, and an extreme CME. As in \citeA{Kay20}, which explored the sensitivity of the original ANTEATR to various input parameters, we expect to see different behavior for a CME that propagates at roughly the background solar wind speed as opposed to significantly faster than it. The initial properties for each CME are listed in Table 1. The mass, velocity, size, and magnetic and thermal scalings all increase with CME scale. 

For all cases, we assume that the CME begins at 10 $R_s$ and stops at L1 (213 $R_s$). We use a $C$ of 1.927 and a $\tau$ of 1 in the magnetic field model, which most closely mimics the Lundquist flux rope model that has been traditionally used to reconstruct in situ CMEs \cite{Lep90}. We assume solar wind values of a density of 6.9 cm$^{-3}$, velocity of 440 km/s, and magnetic field strength of 5.7 nT, which are based on OMNI database averages and the same as used in \citeA{Kay20}. An exploration of the sensitivity to the CME, magnetic field model, and solar wind inputs will be presented in Paper II.

\begin{table}\label{iParams}
\caption{Input Parameters for Different Scale CMEs}
\centering
\begin{tabular}{r c c c}
\hline
  & Average & Fast & Extreme \\
\hline
  $v_{F}$ (km/s)& 600  & 1250 & 2000  \\
  $M_{CME}$ (10$^{15}$ g) & 5 & 10 & 50 \\
  $AW$ ($^{\circ}$) & 30 & 45 & 60  \\
  $AW_{\perp}$ ($^{\circ}$)  & 10 & 15 & 20   \\
  $\delta_{CS}$  & 0.7  & 0.7 & 0.7 \\
  $\delta_{Ax}$   & 0.7 & 0.7 & 0.7  \\
  $\beta_B$  &  2 & 8 & 14 \\
  $\beta_T$  &  2 & 4 & 6 \\
\hline
\end{tabular}
\end{table}

We note that our model requires 14 input parameters (6 CME, 3 flux rope, 1 thermal, and 4 solar wind) plus the IVD and adiabatic index, which may seem daunting for potential space weather applications. We emphasize that the model is entirely specified by these 14 single numbers, whereas more complex models typically require values over the full three-dimensional space and more simplified models contain so many assumptions that some of the underlying physics must be lost. We seek to find a balance between these two extremes, a unique regime not yet explored by many models. Here we show the results that the model is capable of generating on the time scales needed for forecasting, but it remains to be shown that we could feasibly determine every input needed in a real time scenario. We have identified ``reasonable'' default values for a majority of the input parameters and believe we can achieve worthwhile simulations scaling these from common observables such as the CME velocity and full angular width. Paper II shows the sensitivity of each output to the individual inputs and future work will determine the best course for transitioning the model from research into an operation tool, including coupling it to be driven by the ForeCAT coronal CME model.

\section{Ensemble Study Results}
\subsection{CME Size}
\begin{figure}
 \noindent\includegraphics[width=\textwidth]{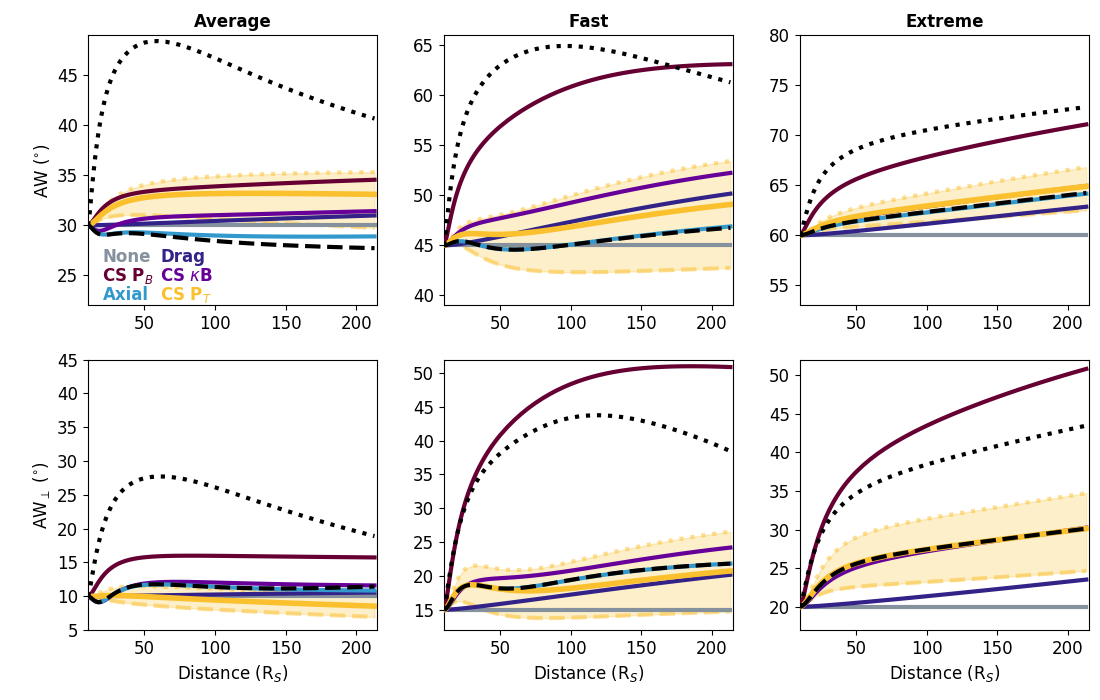}
\caption{Change in $AW$ (top) and $AW_{\perp}$ (bottom) during propagation from 10 $R_S$ to 1 AU. The gray case shows results with no forces included then forces are progressively added and results show in different colors (legend order is left to right, top to bottom.) The yellow shaded region shows the effects of the IVD with the dashed line showing the fully self-similar limit and the dotted line showing fully convective. The columns show results for the average (left), fast (middle), and extreme (right) scale CMEs.}
\label{AW}
\end{figure}

Figure \ref{AW} shows the evolution of the angular width with distance for the ANTEATR-PARADE results. From left to right the columns show the average, fast, and extreme results. The top row shows the full angular width and the bottom row shows the perpendicular angular width. The color indicates which forces are included in each run, with forces being progressively added to the prior set up so that each case includes one new force and all the forces from before. We start with no forces (gray), then add, in order, drag (dark blue), CS magnetic pressure (maroon), CS magnetic tension (purple), axial magnetic forces (light blue), and CS thermal pressure (yellow). The yellow lines represent the inclusion of all forces and for these we also show the self-similar (yellow dashed line) and convective (yellow dotted line) cases and shade the region between, illustrating the effects of the choice of IVD. The black dashed line shows results with fully isothermal expansion and the black dotted line shows fully adiabatic expansion. The final temperatures vary wildly with different expansions so for these cases, we pick the initial thermal scaling $\beta_T$ so that the final temperature is as close as possible to the final temperature of the control case that is halfway between the two extremes. For the isothermal cases this corresponds to $\beta_T$ of 0.25 or less. For the adiabatic, the temperature decreases rapidly due to the increase in volume, so reaching the same final temperature requires $\beta_T$ greater than 200, which is a higher initial temperature than likely realistic. For the slow CME an initial $\beta_T$ greater than 36 causes instabilities in the model due to over expansion so we cannot achieve the same final temperature as the control case.

We first consider the full angular width. With no forces, $AW$ remains constant since that is one of the assumptions made in our IVD. Adding drag increases the transit time of the CME and causes the propagation velocity to decrease relative to the expansion velocities so the $AW$ increases for all scale CMEs. The effect is most noticeable for the fast CME. In general, the effects of drag tend to be most noticeable for the fast CME as the average CME has a smaller velocity differential and the extreme CME has a larger mass.

$AW$ is determined by both the size of the toroidal axis and the CS. Including the CS pressure causes the CS to expand significantly, which also affects the full $AW$. We see that the fast case experiences the most expansion, but this may simply result from our chosen value of $\beta_B$ for each case. When we include the CS tension the CS expansion significantly decreases. For the fast and extreme scales, this case falls between the drag and CS pressure cases. For the average scale, we see a brief contraction when CS tension is included but beyond a few solar radii the $AW$ slightly exceeds that of drag only.

Including the axial forces has a mostly negligible effect on $AW$ for the extreme CME (the purple line falls directly underneath the light blue one). For the average and fast CME we find that the axial forces can cause the CME to alternate between slight contraction or expansion below about 50 $R_S$, but beyond this distance either continued gradual contraction or expansion occurs. Including axial forces leads to smaller $AW$ for both cases.

Next, we include the thermal pressure that drives additional expansion of the CS. This causes a slight increase in the $AW$ relative to the previous case. As the CME expands the temperature and density decrease and eventually the solar wind pressure becomes greater than the CME pressure so that the thermal pressure gradient can slow down the expansion or potentially even cause contraction. We see a slight contraction around 50 $R_S$ for the fast CME but it quickly reverts to gradual expansion. The thermal forces cause an increase in the final $AW$ of a few degrees for the average and fast case but negligible changes for the extreme CME.

The shaded regions show that changes in the IVD can produce changes of 5-10$^{\circ}$ in the final $AW$. A fully convective IVD leads to larger CMEs than a fully self-similar one. The convective IVD causes the CME CS to initially contract in the radial direction, enhancing the internal magnetic and thermal pressure, which leads to additional expansion at farther distances.

Finally, we consider the effects of changes in the adiabatic index. The isothermal case is nearly identical to the case without thermal forces for the fast and extreme CMEs, and only minor differences are visible for the average CME. This is due to our small initial $\beta_T$ values causing the thermal forces to be ineffective near the Sun. With a larger $\beta_T$ we would see a change, but if $\beta_T$ increases significantly the final temperatures will exceed those values typically observed near 1 au. In contrast, changing to fully adiabatic expansion causes significant changes in $AW$. The initial expansion is much more rapid due to the enhanced $\beta_T$. For the average and fast CME contraction begins between 50-100 $R_S$.

The bottom row of Fig. \ref{AW} shows $AW_{\perp}$, the angular width of the CS. In general, the trends are nearly identical to those seen for $AW$.  We see a larger sensitivity to including only CS pressure, but there is little difference between any configurations that includes both components of the magnetic CS forces. The results are again quite sensitive to changing the adiabatic index to fully adiabatic expansion.

\subsection{CME Shape}
\begin{figure}
 \noindent\includegraphics[width=\textwidth]{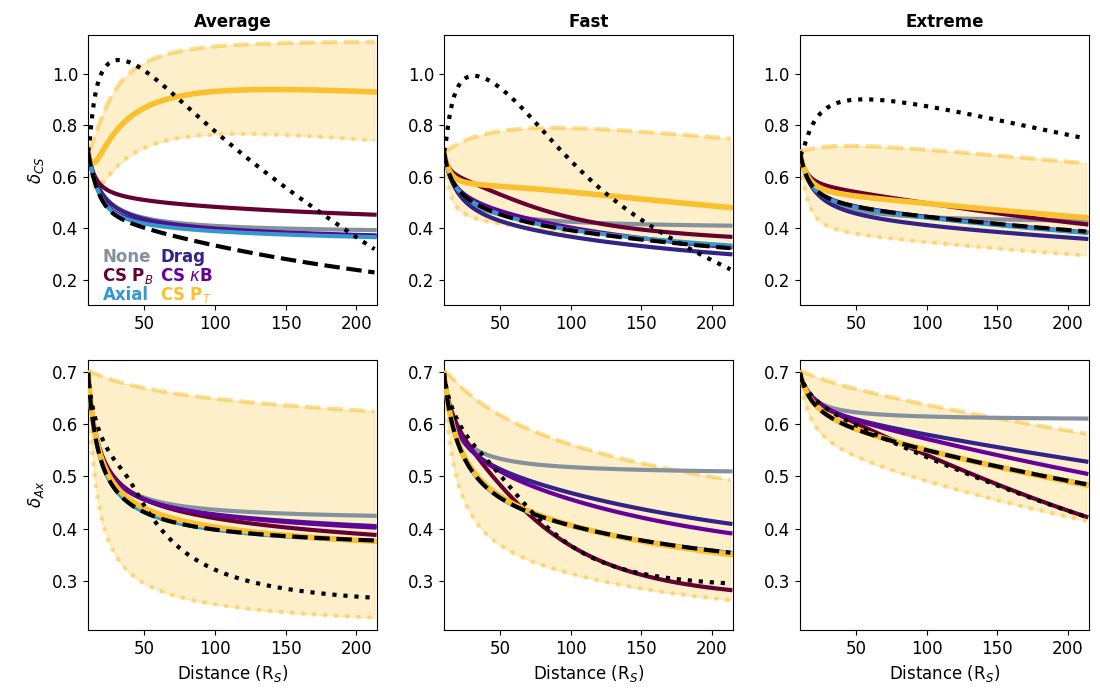}
\caption{Same as Fig. \ref{AW} but for the CS shape (top row) and the axis shape (bottom).}
\label{dels}
\end{figure}

Figure \ref{dels} shows the evolution of the CME shape in the same format as Fig. \ref{AW}. The top row shows $\delta_{CS}$, the ratio of the radial to perpendicular CS widths. With no forces, the change in $\delta_{CS}$ is determined by the IVD. A fully self-similar CME would exhibit no change in $\delta_{CS}$ whereas a fully convective shows a rapid initial decrease followed by a flattening near 50 $R_S$. With no forces and our IVD halfway between the two limits, we see a profile similar to that of the fully convective case but the magnitude of the change is much less. The final $\delta_{CS}$ is about 0.4 for all three scale CMEs without any forces.

Including drag increases the transit time, allowing for further changes from the asymmetries of the IVD, which is most noticeable for the fast CME. The CS magnetic forces produce accelerations that would keep the current $\delta_{CS}$ constant. When combined with the IVD driving toward smaller $\delta_{CS}$ the magnetic forces initially causes the rate of change to decrease. For most cases, we see that the CS magnetic forces cause continued gradual changes instead of stagnating around 50 $R_S$ so that the final $\delta_{CS}$ is equal to or smaller than the drag only case. Axial forces have almost no effect on $\delta_{CS}$ suggesting that there are no secondary effects from change in the axis size and shape on the CS shape.

The thermal forces further negate the changes in $\delta_{CS}$ driven by the IVD. The asymmetry of this force (Eq. \ref{Teq}) depends on the difference between the external thermal pressure around the CS and the CS widths. Both factors cause the force to counteract the effects of the IVD and slow down the changes. For the fast and extreme CME this just further slows the rate of decrease. For the average CME, the thermal forces have a strong effect and $\delta_{CS}$ begins to increase as the CS becomes more circular.

The shaded regions show that the IVD has a very large effect on $\delta_{CS}$. Fully convective cases have much smaller $\delta_{CS}$ due to their smaller initial radial expansion velocities. Changes in the IVD cause changes of order 0.3 in $\delta_{CS}$ for all three cases.

For the changes in the expansion, the isothermal case again mirrors the cases without thermal forces for the fast and extreme CME due to their low $\beta_T$. The isothermal average case shows a continued decrease in $\delta_{CS}$. For all three cases, the fully adiabatic expansion causes $\delta_{CS}$ to quickly increase during the initial phase of rapid expansion. Within 20-30 $R_S$,  $\delta_{CS}$ begins decreasing as the internal pressure drops below that of the external solar wind.

The bottom row of Fig. \ref{dels} shows the change in $\delta_{Ax}$, the ratio of the radial and perpendicular axis widths. Similar to $\delta_{CS}$, a self-similar IVD alone would produce no change in $\delta_{Ax}$, and the convective shows a rapid initial decrease followed by nearly constant behavior. For our force free case we see the convective-like profile but at a smaller magnitude of change. Unlike $\delta_{CS}$, we find that $\delta_{Ax}$ varies with CME scale with the extreme CME showing the smallest change. Including drag causes $\delta_{Ax}$ to continue decreasing out to farther distances since the CME experiences more drag in the radial direction than the perpendicular direction. The fast CME experiences the largest change since it experiences the largest deceleration from drag.

Adding CS pressure causes the decrease in $\delta_{Ax}$ to slow down close to the Sun, but continue at a faster rate farther from the Sun since the larger CS creates more drag. Adding in the magnetic tension produces results with negligibly larger changes than the drag-only case. The axial forces cause an additional decrease in $\delta_{Ax}$ since the axial tension exceeds the hoop forces. This effect is essentially negligible for the average and extreme CME but causes a decrease of 0.05 in the final $\delta_{Ax}$ for the fast CME. We find no noticeable effect from thermal forces, the light blue line falls directly under the yellow line in these figures.

Similar to $\delta_{CS}$, changes in the IVD cause larger variations in $\delta_{Ax}$ than any of the different combinations of forces that include drag. We see variations of 0.2 to 0.4 in $\delta_{Ax}$ with the largest range occurring for the average CME.

The form of expansion has less of an effect on $\delta_{Ax}$ than the IVD. The isothermal cases are nearly identical to the full forces cases for all three scale CMEs. The adiabatic case shows a slower initial decrease in $\delta_{Ax}$ but a more significant gradual decrease over the duration of the propagation, leading to decreases of 0.05-0.15 in the final $\delta_{Ax}$.

\subsection{CME Velocity}
\begin{figure}
 \noindent\includegraphics[width=\textwidth]{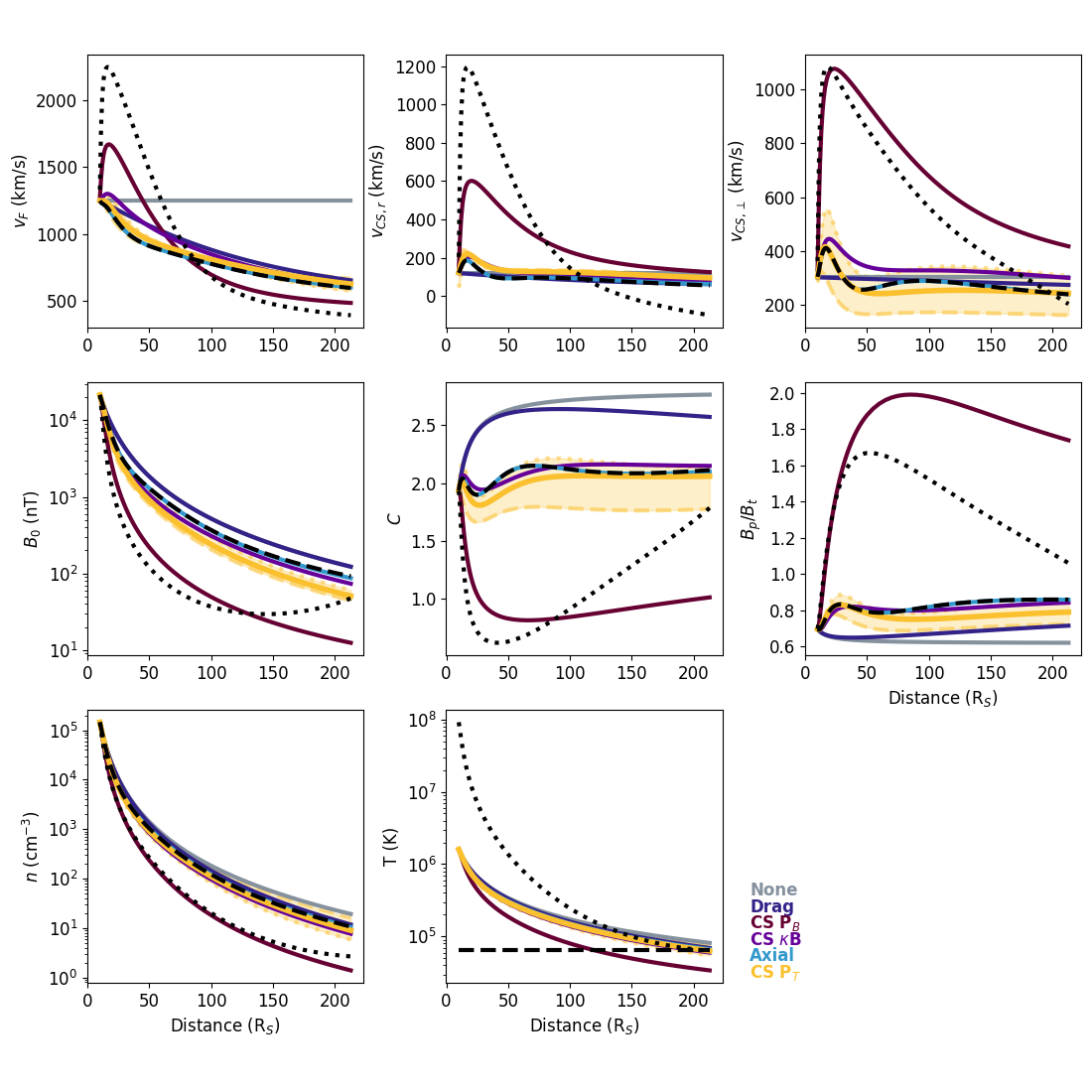}
\caption{Analogous to Fig. \ref{AW}, the change in front velocity and the radial and perpendicular CS velocities (top row left to right), magnetic field strength, the parameter $C$, and the ratio of the poloidal to toroidal magnetic field (middle row), and the number density and temperature (bottom row).}
\label{combo}
\end{figure}

We now consider the evolution of the CME velocity. Figure \ref{combo} is similar to the previous figures but now we only show the fast CME and each panel represents a different output parameter. The general behavior tends to be the same for all scales and we will comment on any difference in the magnitude of the effects where appropriate. The top row shows the propagation and the CS expansion velocities.

The top left panel of Fig. \ref{combo} shows the velocity of the CME front. With no forces the front velocity remains constant and adding drag causes it to continually decrease throughout propagation. The CS magnetic pressure causes $v_F$ to increase since the CS expansion accelerates the front forward initially, but this creates a larger velocity differential with the background solar wind and more drag so that the final $v_F$ is lower than that of the case with only drag. Adding CS tension reduces the initial CS expansion, lowering the peak in $v_F$ and reaching a final value similar to the drag-only case. Adding axis forces causes $v_F$ to initially decrease and thermal forces reduce this decrease. Ultimately, the cases with any magnetic or thermal forces end up with nearly the same final $v_F$. Changes in the IVD cause the final $v_F$ to vary by 100 km/s, about twice the magnitude of variation from different combinations of magnetic or thermal forces. For all panels of Fig. \ref{cartoon}, except for the temperature, the isothermal case shows a negligible difference from the case without thermal forces. For $v_F$, the adiabatic expansion causes a larger initial increase than seen for CS pressure alone, followed by a stronger decrease leading to the lowest final $v_F$.

The top middle panel of Fig. \ref{combo} shows $v_{CS,r}$, the expansion speed of the CS in the radial direction. This is most likely the expansion speed one would infer from an in situ profile assuming an impact near the center of the CS, but the observed value will change by a geometrical factor as the impact moves toward the edge, or decrease if the satellite encounters the CME at an oblique angle such that the expansion is not in the direction of motion. The top right panel shows $v_{CS,\perp}$, the expansion speed of the CS in the perpendicular direction. The variation with different forces follows the same patterns as $v_F$ but both expansion velocities exhibit less sensitivity to the drag due to the smaller velocity differentials in these directions. The perpendicular expansion speed is larger in than the radial one, as expected given that $\delta_{CS}$ is less than one. $v_{CS,\perp}$ also shows a stronger sensitivity to the IVD than $v_{CS,r}$. Fully adiabatic expansion causes a rapid increase in both expansion velocities, followed by a rapid decline. We find that the CME begins contracting in the radial direction beyond about 150 $R_s$.

The average and extreme CMEs show identical trends as the individual forces are progressively added, though the magnitude of the variations changes. The fully-adiabatic, average CME shows the same contraction in the radial direction as the fast CME but the extreme CME continues expanding for the duration of propagation.

\subsection{CME Magnetic Field}
The middle row of Fig. \ref{combo} shows parameters related to the CME's magnetic field. The left panel shows $B_0$, which scales the total magnetic field strength. The total magnetic field strength at the center of the CME equals $\delta_{CS} B_0$ and $B_0$ evolves to maintain conservation of the toroidal magnetic flux. We find that the profiles of $B_0$ with no forces or only drag are visually identical on a log scale. We know that drag delays the CME and allows the CME more time to expand and reduce the total magnetic field strength, but it seems that this change is fully encapsulated by the change in $\delta_{CS}$ rather than the change in $B_0$.

Adding in magnetic pressure causes additional CS expansion, causing $B_0$ to decrease more rapidly. The remaining magnetic force configurations show little difference. Including thermal forces causes a slightly faster decrease over the duration of propagation so that the final $B_0$ decreases from 86 nT to 51 nT. Changes in the IVD can cause $B_0$ to vary by 10 nT with the fully convective case having stronger field than fully self-similar. The case with adiabatic expansion rapidly decreases close to the Sun, but begins increasing by 150 $R_S$ as the CME volume begins decreasing.

The middle panel shows the change in $C$, which scales the poloidal field relative to $B_0$ and evolves based on the relative size and shape of the CME. Since we assume a constant $\tau$ of 1, we expect the flux rope to become kink unstable for a $C$ below 1.7 according to \citeA{Flo20}. The force-free and drag-only cases initially begin increasing. Beyond about 50 $R_s$ the force-free case continues to slowly increase whereas the drag-only starts slowly increasing. For the other force combinations we find a small amount of oscillatory behavior close to the Sun but $C$ remains close to 2 throughout propagation. The IVD causes a range of 0.37 in $C$ with the fully self-similar cases having lower values and dipping into the unstable regime close to the Sun. Adiabatic expansion causes a rapid decrease in $C$ suggesting it is highly unstable and not a realistic option, at least with the chosen input parameters.

The right panel of the middle row shows the ratio of the toroidal magnetic field at the center of the CME and the poloidal magnetic field at the nose of the CME. The ratio of the two is
\begin{equation}
    \frac{B_p}{B_t} = \frac{2}{C (1+\delta_{CS}^2)}
\end{equation}
where we have assumed a $\tau$ of one. The force free case shows a slight decrease in the ratio over the propagation and adding drag causes it to begin slightly increasing by 50 $R_S$. Any CS expansion causes $B_t$ to decrease faster than $B_p$ so including only CS magnetic pressure causes a rapid increase in the ratio. Adding CS tension slows down the CS expansion so that the ratio shows a slight increase, then remains near 0.8 with small oscillations. The behavior does not vary significantly as axial or thermal forces are added and the IVD changes the ratio by 0.1 or less. Adiabatic expansion causes the ratio to initially increase but then it decreases beyond about 50 $R_S$.

For all three magnetic parameters, the average and extreme cases behave largely the same as the fast case. The average CME has a weaker final $B_0$ and the extreme CME a larger one, but this is largely due to our choice of initial parameters. The sensitivity to the thermal forces and form of expansion tends to decrease as the CME scale increases. Conversely, the sensitivity to the IVD increases with scale for both $B_0$ and the ratio.

\subsection{Density and Temperature}
The bottom left panel of Fig. \ref{combo} shows the number density. With no forces, the CME volume stays the smallest and $n$ decreases the least. Drag allows more time for expansion causing a slight decrease from the force free case. Including only CS magnetic pressure causes excessive expansion and a final $n$ of 1.4 cm$^{-3}$. The rest of the forces lead to final number densities of order 10 cm$^{-3}$. The IVD causes changes of about 5 cm$^{-3}$. The adiabatic expansion causes behavior similar to CS magnetic pressure only case for the first 100 $R_S$ but then the adiabatic case has a larger density as the expansion slows.

The middle panel on the bottom of Fig. \ref{combo} shows the temperature. The distinct difference between the purely adiabatic and purely isothermal cases is immediately visible. Full forces leads to a final temperature of 63,000 K and the different force combinations lead to variations of 3,000 K, with the exception of the CS magnetic pressure only, which has a final temperature of 34,000 K. The IVD causes slightly larger changes, of order 10,000 K with the self-similar case having the hottest temperature. For both the number density and temperature we find no significant difference in the trends for the average and extreme CMEs.

\section{Relevance to Space Weather Forecasting}
\begin{figure}
 \noindent\includegraphics[width=\textwidth]{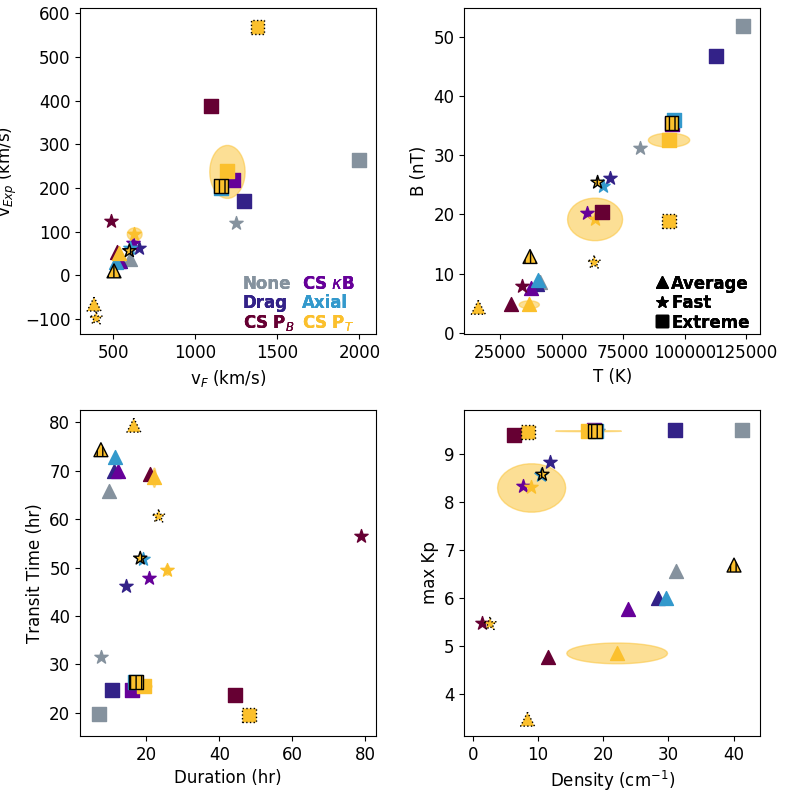}
\caption{Comparison of the 1 AU values of the front and radial expansion velocities (top left), $T$ and $B$ (top right), the duration and transit time (bottom left), and the number density and maximum Kp at the CME front (bottom left).}
\label{dots}
\end{figure}

While the evolution of parameters with distance is interesting and allows us a deeper understanding of the physics involved, the primary benefit of an efficient, simplified model like ANTEATR-PARADE is to eventually provide values relevant to space weather forecasting. Some of these are direct outputs from the model that we have already considered while others are calculated from the outputs. Figure \ref{dots} shows the front and expansion velocity (top left), temperature and magnetic field strength (top right), duration and transit time (bottom left), and number density and estimated maximum $Kp$ (bottom right). All values that vary with distance are taken at the time the CME nose reaches 1 AU. The parameters are grouped based on similarity for display, we are not actively looking for unexpected correlations, rather just the spread in each parameter, but we do see the expected correlations for the velocity pair. Each panel contains the results for the three different scale CMEs using different symbols (triangle is average, star is fast, square is extreme). For each scale CME, the different force configurations are colored the same as in the previous figures. The symbols use the IVD halfway between the two extremes and the yellow-shaded regions represent the variation from the IVD. We determine the maximum change in each parameter from either extreme of the IVD and set that as the length of the shaded ellipse along that axis. We do not have justification for the precise shape of the shaded region, which would require a full ensemble study, so this is simply meant as a visual approximation of the relative importance of the IVD. The yellow symbols with a solid outline and hatching show the isothermal results and those with a dotted outline show the adiabatic results.

The top left panel shows the front velocity and the expansion velocity of the CS in the radial direction. Note the difference in the scales of the axes. The average and fast CMEs occupy fairly similar regions on this scale but are separated by about 100 $km/s$ in $v_F$ and 40 $km/s$. For all scales the largest changes result from using adiabatic expansion, which causes CS contraction for the average and fast CME. The changes from different combinations, assuming CS tension is included, are roughly the came as changes in the IVD.

The top right panel shows the temperature and the magnetic field strength, $B$. For $B$, we show the poloidal magnetic field strength at the front of the CME, which is what would be measured at the time of impact. Here we can better see the spread in these parameters, which was not as visible on the log scale in Fig. \ref{combo}. Both parameters scale with the amount of CS expansion with force combinations leading to more expansion causing weaker values of $T$ and $B$. The results are particularly sensitive to the inclusion of thermal forces. We find some sensitivity to the IVD, which is strongest for the fast CME. Adiabatic expansion leads to weaker magnetic field for the fast and extreme CME. For the adiabatic, average CME, we can not scale the initial temperature to match the final temperature of the baseline case as the forces become unstable. The closest, stable initial temperature leads to roughly the same $B$ but cooler $T$.

The bottom left panel shows the transit time and estimated CME duration, calculated from the CS width and velocities (including expansion) upon impact. We note that these are lower limits on the transit time because we are only considering impacts directly at the nose and the transit time will increase as the impact moves toward the flank. Most cases have a duration between 10-20 hr for all scale CMEs. When there is excessive expansion, from lack of CS tension or the extreme adiabatic expansion, then the duration can increase. When there are no forces the CME propagates the quickest so that the transit time and duration are the shortest. Different force combinations can cause changes of 5-10 hr in the transit time (within each scale). We see comparable change in the transit time from the form of expansion but no visible difference on the IVD.

Finally, the bottom right panel shows the number density and an estimated maximum $Kp$. We calculate the $Kp$ the same as \citeA{Kay20SIT}, which was based on the empirical expression in \citeA{May15AT} 
\begin{equation}\label{eq:Kp}
	Kp = 9.5 - \exp\left[2.17676 - 0.000052v^{4/3} B_{\perp} ^{2/3} \sin^{8/3}\frac{\theta_C}{2}\right]
\end{equation}
Here, $B_{\perp}$ is the transverse component of the magnetic field in Geocentric Solar Magnetospheric coordinates and $\theta_C$ is the clock angle of the magnetic field. We use $B_p$ for $B_{\perp}$ so that we are calculating $Kp$ when the front of the CME first impacts and assume a fully southward clock angle so that this is the maximum expected $Kp$.

For the $Kp$, we see little variation within the extreme scale CMEs because it is so powerful that all Kp values are essentially the maximum possible value of 9.5 from the empirical expression. The fast CME shows variations of about 1 for most changes in the model, but the CS magnetic pressure only or adiabatic case cause Kp to decrease by 4 due to the decreases in both $B$ and $v_F$. The average case shows a wide range in Kp, driven by the different forces or expansion model and not the IVD. 

For all cases we see a large spread in the number density, which results from the variety 
in the CS expansion. The variations for the cases with full CS magnetic forces are less than 5 cm$^{-1}$ but we find a stronger sensitivity to the IVD and expansion model.

\section{Discussion}\label{Disc}
With this work, we have demonstrated the capabilities of ANTEATR-PARADE and shown that it produces reasonable results, but we have not yet validated it at all against observations. Ideally, we would compare the entire profile with values reconstructed from observations from coronagraphs and heliospheric imagers. These observations are readily available and the reconstruction techniques are well-established. We strongly suspect, however, that the uncertainties from the reconstruction techniques are sufficiently large that they would not be of significant use for validation beyond seeming correct in general. Of more use, most likely, will be comparison with the final in situ parameters as we can measure those more directly and more accurately. We have an abundance of 1 AU measurements but comparisons at other distances from either planetary missions such as MESSENGER or Venus Express or with Parker Solar Probe or Solar Orbiter observations would be critical for validating the full model and helping constrain the early CME evolution.

We have picked an order of adding in the forces one by one. We typically find a large change from the effects of drag, then additional changes from adding magnetically-driven CS expansion, but less noticeable changes from axial forces or additional thermally-driven CS expansion/contraction. If we include the thermal expansion before the magnetic forces then we see the large change for the thermal forces and less for the magnetic forces. We generalize this as the effects largely being dominated by drag and some sort of CS expansion. It does not particularly matter if it is only magnetic (constrained by the tension), only thermal, or a combination of the two. These changes in expansion cause some variation in the final parameters, but the effects are much more subtle than whether or not CS expansion is included at all or the effects of drag.

While we have thoroughly investigate the role of different forces, we have only begun to probe the effects of both the IVD and the thermal expansion model. Quite often, models tend to function into ``black or white'' extremes (i.e. adiabatic or isothermal) and while these represent familiar paradigms, the actual processes in nature may fall in gray territory. We have used the midpoint for both the expansion and IVD for our control case and shown that the extremes can produce unrealistic results but have no justification for the midpoints being the correct values. Further work will look at a larger range in these options and comparison with observations (both in general and reproducing specific events) may allow us to say whether CME expansion is more adiabatic-like or isothermal-like or more self-similar-like or convective-like.

We notice that many of the CME properties have two distinct phases with rapid change early on followed by gradual change at farther distances. This is quite similar to the second two phases of the three phase model for a CME's radial propagation in the corona \cite{Zha06, Liu16}. All the forces we consider should be acting upon the CME in the corona, our choice of starting simulations at 10 $R_s$ is somewhat arbitrary. It is simply where we previously started the original ANTEATR simulations, which were designed to follow ForeCAT simulations that ran to 10 $R_s$ because beyond this the external magnetic deflection and rotation forces become negligible. We hypothesize that we may be initiating ANTEATR-PARADE CMEs unnecessarily out of equilibrium and that if we begin simulating the internal magnetic and thermal forces closer to the Sun the rapid change phase may be more concurrent with the rapid radial acceleration phase. We believe that the general results in the paper are still worthwhile, even if the CMEs are initially unbalanced, because they seem to rapidly re-equilibrate within the first 10-20 $R_s$. Right now, we cannot simply start the ANTEATR-PARADE simulations much closer to the Sun as it currently uses a very simplified model of the background solar wind.

We have also only considered a single set of parameters [$m$, $n$] defining our magnetic field model. The chosen pair happen to correspond to magnetic forces that uniformly affect the expansion of the cross section and therefore cannot induce any new asymmetry. Exploration of other pairs of [$m$, $n$] will be critical for determining the extent to which magnetic forces can alter CME cross-sectional shape. 

Future work will incorporate these forces into ForeCAT and we can then use those results to potentially initiate ANTEATR-PARADE simulations more appropriately. Another step is to propagate these advancements in our CME structure and magnetic field into the FIDO model. Adding the elliptical cross section and more flexible magnetic field model into FIDO, the in situ magnetic field model, should help us more accurately reproduce, and eventually predict, the space weather effects of CMEs.

\section{Conclusion}
We present the first results from ANTEATR-PARADE, which uses internal magnetic and thermal and external drag forces to simulate the propagation, expansion, and deformation of a CME in interplanetary space. We analyze the relative contribution of the different forces and find that the results are sensitive to the inclusion of drag and some sort of cross-sectional expansion forces, but less sensitive to the precise nature of this expansion or axial forces, at least for the single parameterization of the magnetic field model used in this work.

We consider two methods for breaking down the total initial speed of a CME front into bulk and expansion components and find that the expansion and deformation are quite sensitive to the chosen values but the propagation less so. We also consider two methods for the thermal expansion - adiabatic and isothermal. Having a fully adiabatic CME match the general temperatures seen near 1 AU requires an initial temperature that exceeds what would be reasonable near 10 $R_s$, which then causes excessive expansion near the Sun, followed by gradual contraction during propagation.  However, fully isothermal expansion and matching 1 AU temperatures requires leads to thermal forces that are essentially negligible in the first 50-100 $R_S$ of propagation. We propose that ANTEATR-PARADE could be particularly useful for helping diagnose the early expansion behavior of CMEs that may be difficult to disentangle in coronal observations. By comparing certain outputs such as the expected in situ expansion velocity, density, or duration, we may be able to constrain an observed CME's initial velocities and adiabatic index.

\acknowledgments
CK is supported by the National Aeronautics and Space Administration under Grant 80NSSC19K0274 issued through the Heliophysics Guest Investigators Program and by the National Aeronautics and Space Administration under Grant 80NSSC19K0909 issued through the Heliophysics Early Career Investigators Program. C. Kay would like to thank S. Kay and D. Kay for their meaningful discussions during this work. The ANTEATR-PARADE code is archived through Zenodo at doi:10.5281/zenodo.4279860

%% ------------------------------------------------------------------------ %%
%% References and Citations

%%%%%%%%%%%%%%%%%%%%%%%%%%%%%%%%%%%%%%%%%%%%%%%
%
% \bibliography{<name of your .bib file>} don't specify the file extension
%
% don't specify bibliographystyle
%%%%%%%%%%%%%%%%%%%%%%%%%%%%%%%%%%%%%%%%%%%%%%%
%\bibliography{master}

%Reference citation instructions and examples:
%
% Please use ONLY \cite and \citeA for reference citations.
% \cite for parenthetical references
% ...as shown in recent studies (Simpson et al., 2019)
% \citeA for in-text citations
% ...Simpson et al. (2019) have shown...
%
%
%...as shown by \citeA{jskilby}.
%...as shown by \citeA{lewin76}, \citeA{carson86}, \citeA{bartoldy02}, and \citeA{rinaldi03}.
%...has been shown \cite{jskilbye}.
%...has been shown \cite{lewin76,carson86,bartoldy02,rinaldi03}.
%... \cite <i.e.>[]{lewin76,carson86,bartoldy02,rinaldi03}.
%...has been shown by \cite <e.g.,>[and others]{lewin76}.
%
% apacite uses < > for prenotes and [ ] for postnotes
% DO NOT use other cite commands (e.g., \citet, \citep, \citeyear, \nocite, \citealp, etc.).
%

\end{document}

%% ------------------------------------------------------------------------ %%
%
%  TITLE
%
%% ------------------------------------------------------------------------ %%

%\includegraphics{agu_pubart-white_reduced.eps}

\title{Supporting Information for "Modeling Interplanetary Expansion and Deformation of CMEs with ANTEATR-PARADE I: Relative Contribution of Different Forces"}
%
% e.g., \title{Supporting Information for "Terrestrial ring current:
% Origin, formation, and decay $\alpha\beta\Gamma\Delta$"}
%
%DOI: 10.1002/%insert paper number here%

%% ------------------------------------------------------------------------ %%
%
%  AUTHORS AND AFFILIATIONS
%
%% ------------------------------------------------------------------------ %%

% List authors by first name or initial followed by last name and
% separated by commas. Use \affil{} to number affiliations, and
% \thanks{} for author notes.
% Additional author notes should be indicated with \thanks{} (for
% example, for current addresses).

% Example: \authors{A. B. Author\affil{1}\thanks{Current address, Antartica}, B. C. Author\affil{2,3}, and D. E.
% Author\affil{3,4}\thanks{Also funded by Monsanto.}}

\authors{C. Kay\affil{1,2}, T. Nieves-Chinchilla\affil{1}}

\affiliation{1}{Heliophysics Science Division, NASA Goddard Space Flight Center, Greenbelt, MD, USA}
\affiliation{2}{Dept. of Physics, The Catholic University of America, Washington DC, USA}

%% ------------------------------------------------------------------------ %%
%
%  BEGIN ARTICLE
%
%% ------------------------------------------------------------------------ %%

% The body of the article must start with a \begin{article} command
%
% \end{article} must follow the references section, before the figures
%  and tables.

\begin{article}

\noindent\textbf{Contents of this file}
%%%Remove or add items as needed%%%
\begin{enumerate}
\item Description of Nonorthogonal Coordinates
\item Generalized Lorentz Force
\item ANTEATR-PARADE Cross Section Forces
\item ANTEATR-PARADE Axial Forces
\end{enumerate}

\noindent\textbf{Introduction}
%Type or paste your text here. The introduction gives a brief overview of the supporting information. You should include information %about as many of the following as possible (when appropriate):
% 1. a general overview of the kind of data files;
% 2. information about when and how the data were collected or created;
% 3. a general description of processing steps used;
% 4. any known imperfections or anomalies in the data.

%\clearpage
The supporting information contains derivations used for the forces within the ANTEATR-PARADE model, the details of which are not essential for understanding the model results, but may be of interest to some readers. The first section briefly summarizes nonorthogonal coordinate systems, which we use for the derivation of a general Lorentz for in elliptic-cylindrical coordinates and the specific forces used in ANTEATR-PARADE in the following sections. 

\noindent\textbf{Nonorthogonal Coordinate Systems} \\
%Type or paste text here. This should be additional explanatory text, such as: extended descriptions of results, full details of models, extended lists of acknowledgements etc.  It should not be additional discussion, analysis, interpretation or critique. It should not be an additional scientific experiment or paper.

As nonorthogonal coordinate systems may be unfamiliar to the reader, we first provide a description of the basics, as applied to the elliptic-cylindrical (EC) coordinate system, to help clarify the derivation of the ANTEATR-PARADE forces. Most of space physics uses orthogonal coordinates with perpendicular axes (e.g. Cartesian, polar, or spherical). In EC coordinates the $r$ and $\phi$ axes are not perpendicular, in general. This introduces complications when expressing vectorial quantities and performing transformations in such non-orthogonal coordinate system.

As in \citeA{Nie18} (hereafter NC18), the transformation between Cartesian coordinates to curved coordinates is given by 
\begin{equation}\label{Rxyz}
    \mathbf{R} = [\delta r \cos \theta, r \sin \theta, z]
\end{equation}
where $\delta$ represents the distortion between the $x$ and $y$ axes. For our model, we apply this to the cross section so $\delta$ and $\theta$ will correspond to $\delta_{CS}$ and $\psi$ but we keep things generic in this section. We can define a set of basis vectors using
\begin{equation}
    \boldsymbol{\epsilon}_i = \frac{\partial\mathbf{R}}{\partial q^i}
\end{equation}
which represent the direction of change for each $q^i$, the contravariant components, but are not necessarily unit vectors. The values that normalize each $\boldsymbol{\epsilon}_i$ are the scale factors $h_i$, which can be found by
\begin{equation}
  h_i = \left| \frac{\partial\mathbf{R}}{\partial q^i} \right|
\end{equation}
and we have the corresponding unit vectors $\mathbf{e}_i = \boldsymbol{\epsilon}_i/h_i$. The projection of a vector, $\mathbf{V}$ onto the $\boldsymbol{\epsilon}_i$ basis gives the nonscaled contravariant components, $V_c^i$, so that $\mathbf{V}=\sum_i V_c^i \boldsymbol{\epsilon}_i$. The scaled contravariant components are defined as $V^i = V_c^i h_i$ such that $\mathbf{V}=\sum_i V^i \mathbf{e}_i$. 

Alternatively, we can form a dual basis of vectors $\boldsymbol{\epsilon}^i$ that are perpendicular to the corresponding $\boldsymbol{\epsilon}_i$. The projection of $\mathbf{V}$ onto $\boldsymbol{\epsilon}^i$ gives the nonscaled covariant components $V_{c,i}$. The scale factors $h^i$ normalize the vectors $\boldsymbol{\epsilon}^i$ and the scaled covariant components are $V_i = V_{c,i} h^i$ and $\mathbf{V}=\sum_i V_i \mathbf{e}^i$. Unlike in orthogonal coordinates, the $\boldsymbol{\epsilon}_i$ are not perpendicular to one another, nor are the $\boldsymbol{\epsilon}^i$. Only an $\boldsymbol{\epsilon}_i$ and an $\boldsymbol{\epsilon}^i$ are perpendicular to one another, and only for the same value of $i$.

We can define the covariant metric tensor with components $g_{ij} = \boldsymbol{\epsilon}_i \cdot \boldsymbol{\epsilon}_j$.  Similarly for the contravariant metric tensor $g^{ij} = \boldsymbol{\epsilon}^i \cdot \boldsymbol{\epsilon}^j$. It can be shown that $g^{ik}g_{kj}=\delta_j^i$, where $\delta_j^i$ is the Kronecker delta and equals 1 if $i$ and $j$ are the same and 0 otherwise.  These tensors can be used to convert between bases.
\begin{eqnarray}
    \boldsymbol{\epsilon}_i = \sum_j g_{ij} \boldsymbol{\epsilon}^j  \\
    \boldsymbol{\epsilon}^i = \sum_j g^{ij} \boldsymbol{\epsilon}_j 
\end{eqnarray}
The same method can be used to convert between covariant and contravariant components of a vector.

For better clarity, we derive the results for EC coordinates, largely following the procedure of \citeA{Nie18}. From Equation \ref{Rxyz}, the first basis is
\begin{eqnarray}\label{basis}
    \boldsymbol{\epsilon}_r = [\delta \cos \theta, \sin \theta, 0] \nonumber \\
     \boldsymbol{\epsilon}_{\theta} = [-r \delta \sin \theta, r\cos \theta, 0] \nonumber \\
      \boldsymbol{\epsilon}_z = [0, 0, 1]
\end{eqnarray}
with corresponding scale factors
\begin{eqnarray}\label{scale}
    h_r = \sqrt{\delta^2\cos^2\theta + \sin^2\theta} \nonumber \\
    h_{\theta} \equiv rh = r \sqrt{\cos^2\theta + \delta^2\sin^2\theta} \nonumber \\
    h_z = 1
\end{eqnarray}
The non-zero components of the covariant metric tensor are
\begin{eqnarray}\label{covG}
    g_{rr} = h_r^2  \nonumber \\
    g_{\theta\theta} = r^2 h^2\nonumber \\
    g_{zz} = 1 \nonumber \\
    g_{r\theta} = g_{\theta r}= r (1-\delta^2) \sin\theta \cos\theta
\end{eqnarray}
which we can use to find the non-zero components of the contravariant metric tensor.
\begin{eqnarray}\label{conG}
    g^{rr} = \frac{h^2}{\delta^2}  \nonumber \\
    g^{\theta\theta} = \frac{h_r^2}{r^2\delta^2} \nonumber \\
    g^{zz} = 1 \nonumber \\
    g^{r\theta} = g^{\theta r}= - \frac{g_{r\theta}}{r^2\delta^2}
\end{eqnarray}

We can determine the dual basis vectors using either the metric tensor (e.g. $ \boldsymbol{\epsilon}^{r} = g^{rr} \boldsymbol{\epsilon}_{r} +g^{r\theta} \boldsymbol{\epsilon}_{\theta} $ or the orthonormality of $\boldsymbol{\epsilon}_{i}$ and $\boldsymbol{\epsilon}^{i}$. For EC coordinates,
\begin{eqnarray}
    \boldsymbol{\epsilon}^{r} = [\frac{1}{\delta} \cos \theta, \sin \theta, 0] \nonumber \\
    \boldsymbol{\epsilon}^{\theta} = \frac{1}{\delta r}[-\sin \theta, \delta \cos \theta, 0] \nonumber \\
    \boldsymbol{\epsilon}^{z} = [0, 0, 1]
\end{eqnarray}

To better understand the meaning of the two different basis, we consider the tangent and normal vectors to an ellipse, $\mathbf{\hat{t}}$ and $\mathbf{\hat{n}}$ respectively, which can be calculated using
\begin{eqnarray}
    \mathbf{\hat{t}} = \frac{\mathbf{R}'}{||\mathbf{R}'||} \\
    \mathbf{\hat{n}} = \frac{\mathbf{\hat{t}}'}{||\mathbf{\hat{t}}'||}    
\end{eqnarray}\label{that}
where $\mathbf{R}$ is the parametric definition of an ellipse, which is the same as Eq. \ref{Rxyz} but with $R$ constant and no $z$ component. The $'$ indicates a first derivative with respect to the parametric $\theta$. We find that $\mathbf{\hat{t}}$ is [$-\delta\sin\theta$, $\cos\theta$]/$h$, which is the same as $\mathbf{e}_{\theta}$ and $\mathbf{\hat{n}}$ is [$-\cos\theta$, -$\delta \sin \theta$]/$h$, which is equivalent to $\mathbf{e}^r$.

Physical intuition for $\mathbf{e}_r$ and $\mathbf{e}^{\theta}$ can be found through polar coordinates. A point in EC coordinates with parametric $\theta$ has a polar angle $\theta_p$ satisfying 
\begin{equation}
   \tan \theta_p = \frac{\sin \theta}{\delta \cos \theta}.
\end{equation}
We call the polar unit vectors $\mathbf{\hat{r}_p}$ and $\mathbf{\hat{\theta}_p}$ to differentiate from the EC coordinates. It can be shown that $\mathbf{\hat{r}_p}$ and $\mathbf{e_r}$ are equivalent, as are $\mathbf{\hat{\theta}_p}$ and $\mathbf{e^{\theta}}$. This means that the first unit basis, the set of $\mathbf{e}_{i}$, corresponds to the polar radial vector, the ellipse tangent vector, and the $z$ direction. The second unit basis, the set of $\mathbf{e}^{i}$, corresponds to the ellipse normal vector, the polar angle vector, and the $z$ direction. The nonscaled $\boldsymbol{\epsilon}$'s incorporate the fact that a uniform rate of change in the parametric $\theta$ does not corresponds to uniform change in Cartesian space.

For our model we need to integrate over the elliptic cross section of a torus. We assume that it is locally described by EC coordinates, that we can ignore any toroidal axis curvature for the small segment we consider. The differential area can be found using the basis vectors for the cross section.
\begin{equation}
    dA = || \boldsymbol{\epsilon}_{\theta} d\theta \times \boldsymbol{\epsilon}_{r} dr || 
\end{equation}
Using Eq. \ref{basis} we find a differential area of $dA = r\delta dr d\theta$, which when integrated from $r$ equals 0 to $R$ and $\theta$ over 2$\pi$ becomes $\delta\pi R^2$, the area of an ellipse.
 \\

\noindent\textbf{Generalized Lorentz Force} \\

To determine a general expression for the Lorentz force in terms of the magnetic field in EC coordinates we begin with Equation 27 from NC18 
\begin{equation}\label{jxB0}
    (\mathbf{j}\times\mathbf{B})_{c,r} = \delta r \left(j_c^y B_c^{\theta} -j_c^{\theta} B_c^y \right)
\end{equation}
which gives the nonscaled covariant component of the force in in terms of the nonscaled contravariant components of the current density, $j$, and magnetic field, $B$. The other covariant components are zero so the total force points in the $\mathbf{e_r}$ or normal direction. The total Lorentz force is split between $r$ and $\theta$ when using contravariant components. 

In EC coordinates, Ampere's law can be expressed as (Eqs. 13 and 14 in NC18)
\begin{eqnarray}
    \partial_{\theta}(g_{r\theta}B_c^{\theta}) - \partial_r(g_{\theta \theta} B_c^{\theta})  &=& 4 \pi \delta r j_c^y \\
    \partial_r(B_c^y) &=& 4 \pi \delta r j_c^{\theta}
\end{eqnarray}
. This allows us to rewrite Eq. \ref{jxB0} as 
\begin{equation}
    (\mathbf{j}\times\mathbf{B})_{c,r} = \frac{1}{4 \pi} \left[ \left(  \partial_{\theta}(g_{r\theta}B_c^{\theta}) - \partial_r(g_{\theta \theta} B_c^{\theta}) \right)B_c^{\theta} -  \partial_r(B_c^y)B_c^y \right]
\end{equation}
. We apply the chain rule to the partial derivatives and rearrange some terms.
\begin{equation}\label{jxblong}
    (\mathbf{j}\times\mathbf{B})_{c,r} = \frac{1}{4 \pi} \left[ g_{r\theta} \partial_{\theta}\left(\frac{(B_c^{\theta})^2}{2}\right) + (B_c^{\theta})^2 \partial_{\theta}\left(g_{r\theta}\right) - g_{\theta \theta} \partial_r \left(\frac{(B_c^{\theta})^2}{2}\right) -B_c^{\theta2}\partial_r(g_{\theta \theta}) - \partial_r\left( \frac{(B_c^{y})^2}{2}\right)\right]
\end{equation}
The derivatives of the metric tensor terms are
\begin{eqnarray}
    \partial_{\theta} (g_{r \theta}) &=& r(1-\delta^2)(\cos^2\theta - \sin^2\theta) \nonumber \\
    \partial_r (g_{\theta \theta}) &=& 2 r (\delta^2 \sin^2 \theta + \cos^2 \theta) = 2 r h^2
\end{eqnarray}
. From Gauss' law $\partial_{\theta}(B_c^{\theta})=0$ so the second term on the right-hand side of Eq. \ref{jxblong} goes to zero. After some algebra we find
\begin{equation}\label{genjxB}
    (\mathbf{j}\times\mathbf{B})_{c,r} = -\frac{1}{4 \pi} \left[ r (1+\delta^2)(B_c^{\theta})^2 + rh \partial_r \left( \frac{(B_c^{\theta})^2}{2}\right) + \partial_r \left( \frac{(B_c^{y})^2}{2}\right) \right]
\end{equation}
where the first term on the right of the equation represents a magnetic tension term and the following two terms are pressure gradient terms. \\

\noindent\textbf{ANTEATR-PARADE Cross Section Forces} \\

To determine the forces that act upon the cross section (CS) in ANTEATR-PARADE, we combine the EC magnetic field from NC18,
\begin{eqnarray}
    B_c^y = \delta_{CS} B_0 (\tau - \bar{r}^2) \\
    B_c^{\psi} = - \frac{2\delta_{CS} B_0}{(1+\delta_{CS}^2)r_{\perp}C} 
\end{eqnarray}
where we have chosen polynomial orders [$m$,$n$] = [0,1], with our expression of the Lorentz force. For this value of $n$, $\partial_{r}(B_c^{\psi})=0$ so there is no pressure gradient contribution from the nonscaled contravariant $B_c^{\psi}$. This will not be true for different values of $n$. Using these expressions for the magnetic field, we find
\begin{equation}\label{jxbBmod}
    (\mathbf{j}\times\mathbf{B})_{c,r} = - \frac{\delta_{CS}^2 B_0^2}{\pi r_{\perp}} \left[\frac{1}{(1+\delta_{CS}^2 C^2} \bar{r} - \frac{1}{2}(\tau \bar{r} - \bar{r}^3) \right]
\end{equation}
where the first term still corresponds to the tension from $B_c^{\psi}$ and the second to the gradient in $(B_c^{y})^2$. We consider a small segment of the torus with width $d\phi$ along the toroidal axis. The volume differential is $dV = \delta_{CS} \bar{r} d\bar{r} r_{\perp}^2 d\psi R_{\kappa} d\phi$, where we have converted our expression for $dA$ using $\bar{r} = r /r_{\perp}$. $R_{\kappa}$ is the radius of curvature of the toroidal axis and we assume that we can treat an infinitely thin segment the same as EC coordinates (i.e. ignore the effects of axial curvature). To get the net force on the thin segment we should integrate over $\bar{r}$ and $\phi$. We want to retain any difference between in force along different $\phi$ to allow the CS shape to change, however, so we only integrate in $\bar{r}$. This gives the nonscaled covariant force.
\begin{equation}\label{Fcr}
    F_c^r = \left(\int_0^1 (\mathbf{j}\times\mathbf{B})_{c,r}   \bar{r} d\bar{r}\right) \delta_{CS} r_{\perp}^2 d\psi R_{\kappa} d\phi = \rho \;  \frac{1}{2} \delta_{CS} r_{\perp}^2 d\psi R_{\kappa} d\phi \; \frac{d^2 \bar{r}}{dt^2}
\end{equation}
and the acceleration of $\bar{r}$. Combining Equations \ref{jxbBmod} and \ref{Fcr} gives
\begin{equation}\label{abar}
    \frac{d^2 \bar{r}}{dt^2} = \frac{\delta_{CS}^2 B_0^2}{\pi \rho r_{\perp}}\left[ \frac{2}{3(1+\delta_{CS}^2)C^2} - \left(\frac{1}{3}\tau - \frac{1}{5} \right)\right].
\end{equation}
We find that Eq. \ref{abar} is not a function of $\phi$ so the nonscaled force is uniform in all directions and $\bar{r}$ will uniformly vary for all $\phi$. If we use $h^r$ to convert it to the scaled value then we find that the acceleration at the nose ($d^2 r_r/dt^2$ at $\phi$=0$^{\circ}$) is $\delta$ times that at the edge ($d r_{\perp}^2/dt^2$ at $\phi$=$\pm$90$^{\circ}$). This means that the Lorentz force can cause the CS to expand or contract but will not produce any distortion in its shape. This will not necessarily hold true for any other choices of [$m$, $n$].
 \\

\noindent\textbf{ANTEATR-PARADE Axial Forces} \\

To calculate the CS forces we assumed we could approximate a thin segment of the torus as locally cylindrical. Now we properly account for the effects of the toroidal axis curvature. This not only introduces a force from the curvature of the toroidal field, but changes the area of the CS through which the poloidal field flows. By flux conservation, this changes the magnitude of the poloidal field and introduces a hoop force from the magnetic pressure gradient.

The differential area for poloidal flux conservation is the length along the toroidal axis, which is local radius of curvature $R_c$ times the differential $d\phi$, multiplied by the radial differential $dr$. The poloidal field, $B_p$, changes as
\begin{equation}
    B_p(r,\psi) = B_{p0} \frac{R_{c0}}{R_c(r,\psi)}
\end{equation}
where $B_{p0}$ is the unaltered poloidal magnetic field from NC18 and $R_{c0}$ is the curvature of the toroidal axis at that $\phi$ (i.e. at the center of the CS).

Accurately determining $R_c$ is difficult because it depends on the curvature of the toroidal axis, which we previously called $R_{\kappa}$, and how this changes as one moves from the axis toward the edge of the CS, which depends on both $r$ and $\psi$. If the toroidal axis is circular then 
\begin{equation}\label{Rc}
    R_c = R_{\kappa} + \delta_{CS} r \cos \psi
\end{equation}
where $r$ varies between $\pm r_{\perp}$, analogous to the method used by \citeA{Wel18} but accounting for the elliptical CS, but this is an oversimplification for the hybrid axis shape. The true value can be found by adding $r \cos\psi \mathbf{\hat{n}}$ to the parametric expression for the axis position ($\mathbf{\hat{n}}$ being the unit normal to the toroidal axis), then calculating $R_c$ using
\begin{equation}
    R_c = \frac{[(x')^2 + (z')^2]^{3/2}}{x'z'' - x''z'}
\end{equation}
where $'$ and $''$ indicate first and second derivatives with respect to $\phi$. We only consider a infinitely thin segment of the torus so we only need to integrate in $r$ (and $\psi$) but not $\phi$. This will still quite quickly become analytically intractable, even if we only consider when $\phi$ equals 0$^{\circ}$ or 90$^{\circ}$ for the nose and flank. Instead, we use the simplification of Eq. \ref{Rc} using the appropriate $R_{\kappa}$ for each $\phi$. For our hybrid shape, $R_{\kappa}$ is $L_{\perp}(2+\sqrt{2})^2/16 \delta_{Ax}$ at the nose and $L_{\perp}(16\delta_{Ax}^2+1)^{3/2}/40 \delta_{Ax}$ at the flank.

The poloidal magnetic field is then
\begin{equation}
    B_p(r,\psi) = B_{p0} \frac{R_{\kappa}}{R_{\kappa}+\delta_{CS}r\cos\psi}
\end{equation}
and the toroidal field, $B_t$ does not change from the version of NC18 because the CS area does not change when axial curvature is included. \\

The force per unit volume from magnetic tension, which points inward along the normal to the toroidal axis is
\begin{equation}\label{tens0}
    f_{\kappa} = \kappa \frac{B_t^2}{4\pi}
\end{equation}
where $\kappa$ is $1/R_c$. We find the total magnetic tension force, $F_{\kappa}$ on our segment by integrating over $r$ and $\psi$ with the volume element $dV = \delta_{CS} r_{\perp}^2 \bar{r} d\bar{r} d\psi (R_{\kappa} + \delta_{CS}r_{\perp}\bar{r}\cos\psi) d\phi$ and using the expression for $B_t$.
\begin{equation}
    F_{\kappa} = \int_0^1 \int_0^{2\pi} \frac{\delta_{CS}^2 B_0^2 (\tau -\bar{r}^2)^2}{4\pi (L_{\perp} + \delta_{CS} r_{\perp} \bar{r} \cos\psi)} \delta_{CS} r_{\perp}^2 \bar{r} d\bar{r} d\psi L_{\perp} (L_{\perp} + \delta_{CS} r_{\perp} \bar{r} \cos \psi) d\phi
\end{equation}
The radius of curvature terms cancel between $\kappa$ and the volume differential and we find 
\begin{equation}
     F_{\kappa} =  \frac{1}{4} \delta_{CS}^3 \left(\tau^2 - \tau + \frac{1}{3} \right)B_0^2 r_{\perp}^2 d\phi = \rho \; \delta_{CS} \pi r_{\perp}^2 R_{\kappa,i}d\phi  \;  \frac{d^2 a_i}{d t^2}
\end{equation}
, which we set equal to density times the volume times the acceleration in the axis length $L_i$ where $i$ is either $r$ or $\perp$.
\begin{equation}
    \frac{d^2 L_i}{d t^2} = \frac{\delta_{CS}B_0^2}{4\pi \rho R_{\kappa,i} } \left( \tau^2 - \tau +\frac{1}{3}\right)
\end{equation}
\\
The hoop force results from the enhancement of the poloidal field on the inward side of the CS relative to the outward side of the CS creating a gradient normal to the toroidal axis. Here we must be careful about proper treatment of the EC coordinates, the axial tension only depends on components for which the scale factor is 1 so we can treat it as a simple orthogonal analysis. For the hoop force, we want the $x$-component of the force from the gradient in the poloidal field. We know the poloidal gradient part of the nonscaled covariant Lorentz force can be expressed as
\begin{equation}\label{fBp}
    \left(f_{\nabla B_{\psi}}\right)_{r,c} = \frac{1}{4\pi}\left[ g_{r\psi} \partial_{\psi}\left(\frac{B_c^{\psi 2}}{2} \right) -  \frac{g_{\psi \psi}}{r_{\perp}} \partial_{\bar{r}} \left(\frac{B_c^{\psi 2}}{2} \right) \right]
\end{equation}
where we have replace the generic EC parametric variable $\theta$ with the $\psi$ specific to our torus CS and used $\bar{r}=r/r_{\perp}$.

The full expression for the poloidal field is
\begin{equation}
    B_c^{\psi} = -\frac{2\delta_{CS}}{1+\delta_{CS}^2} \frac{B_0}{C r_{\perp}} \frac{1}{1+\delta_{CS}\gamma \bar{r} \cos \psi}
\end{equation}
where we have defined $\gamma = r_{\perp} / R_{\kappa}$, which will vary with $\phi$ as $R_{\kappa}$ varies. We calculate the following derivatives for the terms that appear in Eq. \ref{fBp}.
\begin{eqnarray}\label{der}
    \partial_{\psi} \left((1+\delta_{CS} \gamma \bar{r} \cos \psi)^{-2} \right) =  \frac{2\delta_{CS} \gamma \bar{r} \sin \psi}{(1+ \delta_{CS} \gamma \bar{r} \cos \psi)^3}\nonumber \\
    \partial_{\bar{r}} \left((1+\delta_{CS} \gamma \bar{r} \cos \psi)^{-2} \right) = -\frac{2\delta_{CS} \gamma  \cos\psi}{(1+ \delta_{CS} \gamma \bar{r} \cos \psi)^3}
\end{eqnarray}
Substituting values into Eq. \ref{fBp} we find
\begin{multline}
\left(f_{\nabla B_{\psi}}\right)_{r,c} = \frac{\delta_{CS}^2 B_0^2}{2 \pi (1+\delta_{CS}^2)^2 C^2 r_{\perp}^2} \biggl[ r_{\perp} \bar{r} (1 - \delta_{CS}^2) \sin \psi \cos \psi  \left( \frac{2\delta_{CS} \gamma \bar{r} \sin \psi}{(1+ \delta_{CS} \gamma \bar{r} \cos \psi)^3} \right)\\ 
+ r_{\perp} \bar{r}^2 (\delta_{CS}^2 \sin^2 \psi + \cos^2 \psi) \left(\frac{2\delta_{CS} \gamma  \cos\psi}{(1+ \delta_{CS} \gamma \bar{r} \cos \psi)^3}\right) \biggr]
\end{multline}
where we have pulled out the part of  $(B_c^{\psi})^2$ that is not a part of the derivatives in Eq. \ref{der}. Through basic algebra and some trigonometric identities we can reduce this and we use $\boldsymbol{\epsilon}^r \cdot \mathbf{\hat{x}} = \cos \psi / \delta_{CS}$ to get the x component of the force per unit volume.
\begin{equation}\label{fBpx}
    f_{\nabla B_{\psi}, x} = \frac{\delta_{CS}^2 \gamma B_0^2}{\pi (1+\delta_{CS}^2)^2 C^2 r_{\perp}^2} \frac{\bar{r}^2 \cos^2\psi}{(1+ \delta_{CS} \gamma \bar{r} \cos \psi)^3}
\end{equation}
We integrate Eq. \ref{fBpx} over a small segment of the torus as we did for the axial tension force. Pulling constants outside of the integral we have
\begin{equation}
    F_{\nabla B_{\psi}, x} = \frac{\delta_{CS}^3 B_0^2 r_{\perp}^2}{\pi (1+\delta_{CS}^2)^2 C^2} d\phi \; \int_0^1 \int_0^{\2\pi} \frac{\bar{r}^3 \cos^2 \psi}{(1+ \delta_{CS} \gamma \bar{r} \cos \psi)^2} d\psi d\bar{r} 
\end{equation}
. We use the integral
\begin{equation}
    \int_0^{2\pi} \frac{\cos^2 \psi}{(1+A\cos\psi)^2} = \frac{2\pi}{A} \left( 1- \frac{1-2A^2}{(1-A^2)^{3/2}}\right)
\end{equation}
for any constant $A$, which in our case $A=\delta_{CS} \gamma \bar{r}$. This becomes
\\
\begin{equation}
    F_{\nabla B_{\psi}, x} = \frac{2 \delta_{CS} B_0^2 r_{\perp}^2}{\gamma^2 (1+\delta_{CS}^2)^2 C^2} d\phi \; \int_0^1 \left( \bar{r} + \frac{2 \delta_{CS}^2 \gamma^2 \bar{r}^3}{(1-\delta_{CS}^2 \gamma^2 \bar{r}^2)^{3/2}} - \frac{\bar{r}}{(1-\delta_{CS}^2 \gamma^2 \bar{r}^2)^{3/2}} \right) dr
\end{equation}
which ultimately reduces to
\begin{equation}
    F_{\nabla B_{\psi}, x} = \frac{B_0^2 r_{\perp}^2}{C^2} d\phi \frac{\sqrt{1-\delta_{CS}^2\gamma^2} (\delta_{CS}^2 \gamma^2 -6) + 6 - 4 \delta_{CS}^2 \gamma^2}{\delta_{CS} \gamma^2 (1+\delta_{CS}^2)^2  \sqrt{1-\delta_{CS}^2\gamma^2}}
\end{equation}
which we set equal to the volume times the density times an acceleration in the axis length. We find an axial acceleration of 
\begin{equation}
    \frac{d^2L_i}{dt^2} = \frac{B_0^2}{\pi \rho C^2 R_{\kappa,i}} d\phi \frac{\sqrt{1-\delta_{CS}^2\gamma^2} (\delta_{CS}^2 \gamma^2 -6) + 6 - 4 \delta_{CS}^2 \gamma^2}{\delta_{CS}^3 \gamma^2 (1+\delta_{CS}^2)^2  \sqrt{1-\delta_{CS}^2\gamma^2}}
\end{equation}
, which we express as the second derivative of the axial length $L_i$ where $i$ is either $r$ or $\perp$ for the radial or flank and $R_{\kappa,i}$ and $\gamma_i$ are the appropriate values for that location.
%%%%%%%%%%%%%%%%%%%%%%%%%%%%%%%%%%%%%%%%%%%%%%%
% if you get an error about newblock being undefined, uncomment this line:
%\newcommand{\newblock}{}

%\bibliography{master} 

\end{article}
\clearpage